\documentclass[12pt,epsf]{article}

\usepackage{color}
\usepackage{epsfig}
\usepackage{epstopdf}
\usepackage{graphicx}
\usepackage{times}
\usepackage{amsmath}
\usepackage{amssymb}
\usepackage{hyperref}

\newcommand{\be}{\begin{equation}}
\newcommand{\ee}{\end{equation}}
\def\({\left (}
\def\){\right )}
\def\[{\left [}
\def\[{\right ]}

\begin{document}
\begin{titlepage}
\bigskip

\rightline
\bigskip\bigskip\bigskip
\centerline {\Large \bf {Solitons in Five Dimensional Minimal Supergravity:}}
\centerline {\Large \bf {Local Charge, Exotic Ergoregions, and Violations of the BPS Bound}}
\bigskip\bigskip
\bigskip\bigskip

\centerline{\large Geoffrey Comp\`ere${}^1$ ${}^\ast$, Keith Copsey${}^2$${}^{\ddagger}$, Sophie de Buyl${}^1$ ${}^{\star}$, and Robert B. Mann${}^2$ ${}^\dagger$}
\bigskip\bigskip
\centerline{\em ${}^1$Department of Physics, University of California at Santa Barbara, Santa Barbara, CA 93106, USA}
\centerline{\em ${}^2$ Department of Physics, University of Waterloo, Waterloo, Ontario N2L 3G1, Canada}
\bigskip
\centerline{\em ${}^\ast$ gcompere@physics.ucsb.edu, ${}^\ddagger$ kcopsey@scimail.uwaterloo.ca}
\centerline{\em ${}^\star$ sdebuyl@physics.ucsb.edu, ${}^\dagger$ rbmann@sciborg.uwaterloo.ca}
\bigskip\bigskip

\begin{abstract}
We describe a number of striking features of a class of smooth solitons in gauged and ungauged minimal supergravity in five dimensions.  The solitons are globally asymptotically flat or asymptotically AdS without any Kaluza-Klein directions but contain a minimal sphere formed when a cycle pinches off in the interior of the spacetime.   The solutions carry a local magnetic charge and many have rather unusual ergosurfaces.  Perhaps most strikingly, many of the solitons have more electric charge or, in the asymptotically AdS case, more electric charge and angular momentum than is allowed by the usual BPS bound.   We comment on, but do not resolve, the new puzzle this raises for AdS/CFT.
\end{abstract}

\end{titlepage}

\baselineskip=16pt
\setcounter{equation}{0}

\section{Introduction}

Over the last number of years it has become clear that gravity in more than four dimensions has qualitatively different solutions from those in four dimensions, included the celebrated black rings (see \cite{ERBHreview} for a recent review).  While there has been substantial progress in describing all possible black holes, at least in five dimensions, there is still relatively little known about topologically nontrivial stationary regular solutions (i.e. solitons) in higher dimensions.   It is worth noting such solutions are only possible in asymptotically flat or asymptotically AdS spacetimes with more than four dimensions.  

In four dimensions there are two key sets of results that forbid such objects.  The first is a result due to Gannon \cite{Gannon} that shows, provided the weak energy condition is obeyed, spacetimes that are not simply connected necessarily contain singularities (i.e. are geodesically incomplete).  This does not stop one from writing down non-simply connected smooth initial data, but if one does so the evolution of the spacetime is guaranteed to produce a singularity.  The second set  of results go under the name of topological censorship \cite{topcensor} and demonstrate that, given the averaged null energy condition, all causal curves going from past null infinity to future null infinity are homotopic to topologically trivial curves--in other words any nontrivial topology is always invisible from infinity (e.g. hidden behind a horizon).

In four dimensional asymptotically flat or asymptotically AdS spacetimes if one pinches off a cycle to produce a nontrivial topology, the resulting minimal surface is a circle and one runs afoul of the above results.  However, one can smoothly pinch  off a circle in five dimensions so the minimal surface is (topologically) an $S_2$, thereby avoiding the above obstructions.  Such spacetimes are sometimes referred to as ``bubbles of nothing'', as there is no spacetime inside these ``bubbles''.  Note, however, they need not, and in the present context will not, require an asymptotic Kaluza-Klein direction.   Hence, we avoid the usual problems of instabilities and the violation of supersymmetric boundary conditions associated with Kaluza-Klein bubbles \cite{Wittenbubble}.

A purely gravitational static soliton of nearly the type we seek is known in five dimensions \cite{EH}.  This (locally) asymptotically flat solution, dubbed the Eguchi-Hanson soliton, is formed by writing $S_3$ via the Hopf fibration and pinching off the $S_1$ of that fibration in the interior of the spacetime.   As we will later briefly review, one can show generically \cite{GibbonsSquashed}, at least in the absence of ergoregions and cosmological constant, that there are no globally asymptotically flat smooth solitons.  One must either allow for a conical singularity at the bubble or quotient the asymptotic $S_3$.  Intuitively, gravity tends to make the bubble collapse and the quotient or conical singularity reflect the fact that without matter or angular momentum there is nothing to hold the bubble up.  A class of solution generating techniques adding flux and angular momenta in the asymptotically flat case is known (\cite{Mizoguchi:1998wv}-\cite{GSG2}) and one might wonder if one could use these methods to eliminate the conical singularity in the globally asymptotically flat Eguchi-Hanson soliton.

The answer turns out to be in the affirmative.  Furthermore, this generated solution ends up being a small subset of the broad class of solutions one can construct, in both asymptotically flat and asymptotically AdS spacetimes, by considering the black hole solutions  of \cite{CCLP05} and choosing values of the parameters such that a cycle is smoothly pinched off in the interior of the spacetime and no horizon is present.  This approach to the solutions of \cite{CCLP05} has previously been examined by Ross \cite{RossBubbles} and part of our work will reproduce that analysis.  However there are several key differences from, as well as some generalizations of, that work in our analysis.  We also discuss a variety of striking features that were not discussed in \cite{RossBubbles} or, as far as we know, anywhere else.  In particular, some of the properties of the asymptotically AdS solutions present some intriguing new questions in relation to the AdS/CFT correspondence.

We begin by reviewing the solution of \cite{CCLP05} and then show how to pinch off a cycle to form the desired solitons.  We show one may choose parameters such that the solutions are smooth and closed timelike curves are not present.  All such smooth solutions turn out to have a nonzero local magnetic charge that may be defined by integrating a flux over the minimal $S_2$ of these solitons.   After constructing the asymptotically conserved charges in the fourth section, we consider the detailed properties of various classes of these solitons, paying particular attention to ergoregions and the BPS bound.  In particular, we find asymptotically flat solutions that variously satisfy, saturate, and violate the BPS bound (i.e. have more electric charge than mass).  All of the asymptotically AdS solitons we have found violate the standard BPS bound.   All the solitons we find, with the exception of a few small subclasses, possess ergoregions. In no case do we find only a single ergosphere surrounding the entire bubble.  Rather we find a variety of unusual ergosurfaces including inner and outer ergospheres that surround the bubble, as well as ergosurfaces we will refer to as capping spheres, which are topologically spheres but run into the bubble.  We finally end with some comments focusing on the violation of the BPS bound and the puzzle these objects pose for AdS/CFT.

\setcounter{equation}{0}
\section{The Chong-Cvetic-Lu-Pope solution}

We will consider solutions in gauged and ungauged minimal supergravity
\be \label{Sact}
S = \frac{1}{16 \pi G} \int d^5 x \sqrt{-g} (R +12 g^2 - \frac{F_{a b} F^{a b}}{4} - \frac{1}{12 \sqrt{3}}  \epsilon^{a b c d e} F_{a b} F_{c d} A_e)
\ee
  In the case where $g \neq 0$, the AdS length $l$ is given by $g = 1/l$.  A broad class of solutions to this theory, including all topologically spherical black holes with two commuting axial isometries \cite{TYI09}, was given by  Chong et al \cite{CCLP05}
$$
ds^2 = -\frac{\Delta_\theta [ (1+g^2 r^2) \rho^2 dt + 2 q \nu] dt}{\Xi_a \Xi_b \rho^2} + \frac{2 q \nu \omega}{\rho^2} + \frac{f}{\rho^4} \Big( \frac{\Delta_\theta dt}{\Xi_a \Xi_b} - \omega \Big)^2
$$
\be \label{Chong1}
+ \frac{\rho^2 r^2 dr^2}{W} + \frac{\rho^2 d\theta^2}{\Delta_\theta} + \frac{r^2+a^2}{\Xi_a} \sin^2 \theta d \phi_1^2 + \frac{r^2+b^2}{\Xi_b} \cos^2 \theta d \phi_2^2
\ee
and
\be \label{potorig}
A = s_{\epsilon} \frac{\sqrt{3} q}{\rho^2} \Big( \frac{\Delta_\theta }{\Xi_a \Xi_b} dt - \omega \Big) + K_0 d \phi_1 + K_1 d \phi_2
\ee
where
\be
\nu = b \sin^2 \theta d \phi_1 + a \cos^2 \theta d \phi_2
\ee
\be
\omega = \frac{a}{\Xi_a} \sin^2 \theta d \phi_1 + \frac{b}{\Xi_b} \cos^2 \theta d \phi_2
\ee
\be
\Delta_{\theta} = 1 - a^2 g^2 \cos^2 \theta - b^2 g^2 \sin^2 \theta
\ee
\be
W = (r^2 + a^2) (r^2+b^2) (1 + g^2 r^2) + q^2 + 2 a b q - 2 m r^2
\ee
\be
\rho^2 = r^2 + a^2 \cos^2 \theta + b^2 \sin^2 \theta
\ee
\be
f = 2 m \rho^2 - q^2 + 2 a b q g^2 \rho^2
\ee
\be
\Xi_a = 1 - a^2 g^2 \, \, , \, \, \, \, \, \, \, \, \, \, \, \Xi_b = 1 - b^2 g^2 
\ee
and
\be
s_{\epsilon} = \pm 1
\ee
where both signs yield solutions provided one takes the orientation of the spacetime to be
\be
\epsilon_{t \phi_1 \phi_2 \theta r} = s_{\epsilon} \sqrt{-g}
\ee
Note the sign of the gauge potential and the sign of the orientation are correlated\footnote{ It is possible to absorb this sign into a redefinition of the Chern-Simons coupling in (\ref{Sact}), effectively changing the theory.
We shall adopt the choice given in equation (\ref{Sact}).  }.

We have made some minor notation changes from \cite{CCLP05} for later convenience and generalized the gauge potential with an additional overall sign and extra constants $K_0$ and $K_1$.   The extra sign $s_{\epsilon}$ will allow us to explicitly keep track of the choice of orientation of the spacetime and will ultimately show up in the sign of the electric charge.  Aside from places where $\phi_1$ and $\phi_2$ degenerate, adding the constants $K_0$ and $K_1$ is just a gauge transformation.   It will turn out that when we tune parameters of the above solution such that we can find smooth solitons there are additional places where $\phi_1$ and $\phi_2$ degenerate besides $\theta = 0$ and $\theta = \pi/2$ and if $K_0$ and $K_1$ are not chosen properly one does not obtain entirely smooth solutions.

While in the case $g = 0$ the metric is manifestly asymptotically flat, for $g \neq 0 $ the solution does not look asymptotically AdS.  This is, however, merely due to the choice of coordinates and by defining
\be
\bar{r}^2 = \frac{r^2 \Delta_{\theta} +  a^2 \sin^2{\theta} + b^2\cos^2\theta - a^2 b^2 g^2}{\Xi_a \Xi_b}
\ee
and
\be
\cos(2 \bar{\theta}) = \frac{ r^2 [ \Xi_a \cos^2\theta  - \Xi_b \sin^2{\theta} ] - a^2 \Xi_b \sin^2{\theta}  + b^2 \Xi_a \cos^2\theta }{ r^2 [ \Xi_a \cos^2\theta  + \Xi_b \sin^2{\theta} ] + a^2 \Xi_b \sin^2{\theta}  + b^2 \Xi_a \cos^2\theta }
\ee
the metric in terms of $(\bar{r}, \bar{\theta})$ is manifestly globally asymptotically AdS.

Now let us consider the regularity of this solution.  The necessary and sufficient condition for the absence of closed timelike curves (CTCs) at infinity is that  $\Xi_a >0$ and $\Xi_b >0$, or equivalently that  $\vert ag  \vert < 1$ and $\vert b g \vert < 1$.   The radial component of the metric will diverge if $W(r)$ has a zero, but this is merely a coordinate singularity and easily removed.  The metric components are otherwise divergence free unless there is some locus where $\rho^2 = r^2 + a^2 \cos^2 \theta + b^2 \sin^2 \theta= 0$.  At first glance one might think this would never occur,at least if $a$ and $b$ are nonzero, but if one tried to restrict $r$ to only real values one would not obtain a geodesically complete spacetime.  In these coordinates $r = 0$  does not correspond to a place where a sphere or other cycle has zero volume.   For small $r$, $g_{r r} \sim r^2$.   One can remove the coordinate singularity at $r = 0$ by defining $R = r^2$ and then it is straightforward to check that the spacetime and its geodesics continue through to $R < 0$ (or imaginary r).  If the spacetime continues to negative enough $R$, $\rho$ will go through a zero.  It is straightforward to check for $q \neq 0$ the Ricci scalar diverges if $\rho$ vanishes and this is a physical singularity.  So if we want to find a totally regular solution (with $q \neq 0$) we must not allow $r^2$ to become so negative  that $\rho$ ever vanishes.   This then implies
\be \label{condit1}
r^2 > - a^2
\ee
if $b^2 \geq a^2$ and
\be \label{condit2}
r^2  > - b^2
\ee
if $a^2 \geq b^2$.

The determinant of the solution is given by the remarkably simple expression
\be
-\frac{r^2 \rho^4 \sin^2 \theta \, \cos^2 \theta }{(1 - a^2 g^2)^2 (1 - b^2 g^2)^2}
\ee
and so, aside from the coordinate singularity due to $g_{r r}$ at $r = 0$ (which, as noted above, may be removed by going to a new radial variable $R = r^2$) and the usual coordinate singularities at the poles of the $S_3$ at $\theta = 0$ and $\theta = \pi/2$, the metric is invertible provided $\rho > 0$.   Note at first glance one might worry the metric changes signature when we continue to imaginary $r$ but this is merely an artifact of using bad coordinates;  if one uses the radial coordinate $R$ the factor $r^2$ is removed from the determinant and the signature is negative definite everywhere.

\setcounter{equation}{0}
\section{Finding solitons}

\subsection{Choosing a cycle}

As we saw above, the solution we begin with encounters a singularity in the deep interior.  For black hole solutions, provided this singularity is behind a horizon, this does not concern us any more than any other black hole singularity.  In the absence of a horizon the only way to avoid this singularity is to keep geodesics from extending this far.   We will do this by smoothly pinching off a periodic direction, leaving the spacetime with a minimal surface, or bubble.  The result will ultimately be a geodesically complete spacetime that is entirely regular (or at worst has orbifold singularities) and hence is necessarily horizon free.

If one tries to pinch off the directions $\phi_1$ or $\phi_2$ one generically only finds singular solutions, essentially due to the fact there are places on the $S_3$ where these directions already vanish.   Instead we may define a new periodic direction $\psi$ (which will pinch off in the interior of the spacetime) and a second direction $\phi$ as linear combinations of $\phi_1$ and $\phi_2$
\be
\frac{\partial}{\partial \psi} = \alpha \frac{\partial}{\partial \phi_1}+ \gamma \frac{\partial}{\partial \phi_2}
\ee
\be
\frac{\partial}{\partial \phi} = \beta \frac{\partial}{\partial \phi_1}+ \delta \frac{\partial}{\partial \phi_2}
\ee
In terms of coordinates this implies
\be \label{ang1}
\phi_1 = \alpha \psi + \beta \phi
\ee
\be \label{ang2}
\phi_2 = \gamma \psi + \delta \phi
\ee
In order for the transformation to be invertible we must have $\alpha \delta - \beta \gamma \neq 0$.   One can show that if one tries to pinch off the $\partial/{\partial \psi}$ with $\alpha = 0$ one necessarily encounters one of the singularities we will describe latter in detail, so we restrict ourselves to $\alpha \neq 0$.  Rescaling the angles $\psi$ and $\phi$ allows one to absorb two of the four parameters in the above (\ref{ang1}, \ref{ang2}), say $\alpha$ and $\delta$ (or, in the case $\delta$ vanishes, $\alpha$ and $\beta$).  The ratio $\gamma/\alpha$ parametrizes the particular cycle we are pinching off and, as we will see below, its value will influence the physical properties of our solitons.  Presuming $\delta \neq 0$, the remaining parameter $\beta/\delta$ parametrizes the remaining freedom in choosing a particular $\partial/{\partial \phi}$; there does not appear to be any obstruction to defining a global gauge transformation to set $\beta/\delta$ to any desired value and hence regarding it as pure gauge.

We wish to insist $\psi$ is a periodic coordinate, so that we may pinch it off in the interior of the spacetime.  This then forces $\gamma/\alpha$ to be rational and without loss of generality we may take
\be \label{gammarat1}
\frac{\gamma}{\alpha} = \frac{m_1}{m_0}
\ee
where $m_1$ and $m_0$ are relatively prime integers and $m_0 > 0$.  $\phi$, on the other hand need not be periodic, and indeed in the case of the Hopf fibration one can only make $\phi$ periodic at the cost of quotienting the asymptotic sphere  (see Appendix I).  Instead we choose to construct solutions which are globally asymptotically flat or globally asymptotically AdS.  Following through a similar analysis (Appendix II) as that for the Hopf fibration, we find the period of $\psi$ must be
\be \label{Dpsicondit}
\Delta \psi = \frac{2 \pi m_0}{\vert \alpha \vert}
\ee
and the range of $\phi$ must be
\be
\Delta \phi = \frac{2 \pi}{ \vert m_0 \delta - m_1 \beta \vert}
\ee
As we explain in detail in the appendix, while away from the bubble $\phi$ is generically better described as an azimuthal angle rather than a polar one, due to the degeneration of $\partial/{\partial \psi}$ at the bubble surface it becomes effectively periodic there and we will be left with an $S_2$ parametrized by $\theta$ and $\phi$.

In these new coordinates it will be handy to define
\be
C_0 = \alpha K_0 + \gamma K_1
\ee
and
\be
C_1 = \beta K_0 + \delta K_1
\ee
so that the potential is
\be \label{newpot}
A = \frac{\sqrt{3} q}{\rho^2} \Big( \frac{\Delta_\theta }{\Xi_a \Xi_b} dt - \omega \Big) + C_0 d \psi + C_1 d \phi
\ee

\subsection{Pinching off a cycle}

 We wish to pinch off the $\psi$ direction to form a soliton, producing a minimal surface which is topologically an $S_2$, parametrized by $(\theta, \phi)$.    The only way we can do this is if there is some surface upon which $g_{t \psi} = g_{\psi \psi} = g_{\psi \phi} = 0$.   The simplest of these conditions is $g_{t \psi} = 0$.  Defining  as before $R = r^2$, we can then solve
\be
g_{t \psi} (R_0) = 0
\ee
for $R_0$.  Note at this stage $R_0$ is not a constant but a function of $\theta$.  As we will discuss in detail below, we will only be able to find an entirely smooth soliton if we can choose $C_0$ such that $A_{\psi}$ vanishes at $R = R_0$.   Hence we must demand that $A_{\psi} (R_0)$ is a constant.  Finally insisting that  $g_{\psi \psi} (R_0) = 0$, straightforward, if slightly tedious, algebra shows we have a chance at smooth solitons only if $a$ and $b$ are nonvanishing and
\be \label{condit3}
 \frac{\gamma}{\alpha} = \frac{m_1}{m_0} = \frac{a(1 - b^2 g^2) [a^2 - b^2 + s_0 \sqrt{(a^2-b^2)^2-4 a b q}]}{b (1 - a^2 g^2) [b^2 - a^2 + s_0  \sqrt{(a^2-b^2)^2-4 a b q]}}
\ee
\be \label{condit4}
m = - \frac{q}{4 a b} \Big[ a^2 + b^2 + 2 a^2 b^2 g^2 + s_0 \sqrt{ (a^2 - b^2)^2 - 4 a b q} \Big]
\ee
and
\be \label{bubblerad}
R_0 = \frac{1}{2} \Big[ - (a^2+b^2) + s_0 \sqrt{ (a^2-b^2)^2 - 4 a b q} \Big]
\ee
where $s_0 = \pm 1$.  Further with the above one finds $g_{\psi \phi} (R_0) = 0$ and so we have accomplished our goal of pinching off the $\psi$ direction.   One also automatically finds $W(R_0) = 0$, but as noted before this is merely a coordinate singularity.

 We will turn in a moment to the question of smoothness of the manifold near $R = R_0$ but first let us note that $\rho^2$ is positive definite for $r^2 \geq R_0$ if and only if
\be \label{condit5}
s_0 =1
\ee
and
\be \label{condit6}
a b q < 0
\ee
and since we are interested in smooth solutions we henceforth adopt (\ref{condit5}) and (\ref{condit6}).  Note that (\ref{condit6}) also automatically assures the reality of (\ref{condit3} - \ref{bubblerad}).
In order for $R$ to remain a spacelike direction we must ensure $W(R)$ has no zeroes for $R > R_0$ and only a simple zero at $R_0$; this will follow if and only if
\be \label{condit7}
0 < - a b q < a^2 b^2
\ee
Note this will then imply that $R_0 < 0$, although this, again, is simply due to the choice of coordinates used to write down the solution and does not reflect any pathology in the spacetime.  It is convenient to define
\be
Q = -\frac{q}{a b}
\ee
so that (\ref{condit7}) becomes
\be \label{condit8}
0 < Q < 1
\ee

We note, incidentally, we do not find any purely gravitational smooth solutions (i.e. $q = 0$).  Intuitively, gravitation would like to make the would-be soliton collapse and one needs flux to stabilize the solution.  As mentioned before, in the absence of a cosmological constant one can prove generically there are no globally asymptotically flat purely gravitational solitons \cite{GibbonsSquashed}, at least provided one assumes the absence of an ergoregion.   Given the vacuum einstein equations and a familiar maximization argument, one can show such would-be solutions are necessarily the product of a flat time direction and a Riemannian Ricci flat manifold.  The existence of solitons is then equivalent to the existence of nontrivial asymptotically Euclidean solutions and since there are none \cite{Witten1981mf}, there are no globally asymptotically flat solitons.  We are not aware, however, of a generalization to nonzero cosmological constant. 

\subsection{Removing conical singularities}

Now turning to the issue of smoothness near $r^2 = R = R_0$, if we define
\be
z = \sqrt {R - R_0}
\ee
then the worrisome part of the metric is
\be \label{conicmet}
ds^2 = \frac{\rho^2}{K_2} (d z^2 + K_2 K_3  z^2 d\psi^2) + a_0 (\theta) z^2 dt d\psi + a_1 (\theta) z^2 d\psi d\phi + \ldots
\ee
where the constants $K_2$ and $K_3$ are given by
\be
K_2 = -\frac{R_0}{2 a^2 b^2} \Big[ (a^2-b^2)^2+2 a^2 b^2 Q + (a^2 \Xi_b + b^2 \Xi_a) \sqrt{(a^2-b^2)^2+ 4 a^2 b^2 Q} \Big]
\ee
\be
K_3 = \frac{ 2 \alpha^2 \Big[ (a^2 - b^2)^2 + 2 a^2 b^2 Q + (a^2 \Xi_b + b^2 \Xi_a) \sqrt{(a^2-b^2)^2+4 a^2 b^2 Q} 
\Big]}{b^2 (1 - a^2 g^2)^2 \Big[b^2 - a^2 +  \sqrt{(a^2-b^2)^2 + 4 a^2 b^2 Q} \Big]^2}
\ee
The omitted terms in (\ref{conicmet}) are generically nonvanishing, except at the axis $\theta = 0$ and $\theta = \pi/2$; these are simply the poles of the $S_2$ of the bubble and will be discussed below.  
Given (\ref{condit8}), both $K_2$ and $K_3$ are positive definite.  We will not be concerned with the precise forms of $a_0(\theta)$ and $a_1(\theta)$, although of course they may be calculated using the above.   At worst the only lack of smoothness will be a conical singularity.   Generically we may wish to allow a $Z_k$ orbifold singularity at the bubble.   Then we must impose
\be
\sqrt{K_2 K_3} = \frac{\vert \alpha \vert}{m_0 k}
\ee
or equivalently
\be \label{conic1}
- \frac{ R_0 \Big[ (a^2-b^2)^2 + 2 a^2 b^2 Q + (a^2 \Xi_b + b^2 \Xi_a) \sqrt{(a^2-b^2)^2 + 4 a^2 b^2 Q} \Big]^2}{a^2 b^4 \Xi_a^2 \Big[b^2 - a^2 + \sqrt{(a^2-b^2)^2 + 4 a^2 b^2 Q} \Big]^2} = \frac{1}{m_0^2 k^2}
\ee

The above does not quite ensure the absence of all conical singularities.  As we remarked above, the $\phi$ direction becomes periodic at the bubble.  One can check $\phi$ degenerates at the poles of the remaining $S_2$--namely $\theta = 0$ and $\theta = \pi/2$.  These are simply the places where the axis $\partial/{\partial \phi_1}$ and $\partial/{\partial \phi_2}$ run into the bubble.  Away from the bubble, our coordinates $\psi$ and $\phi$ are equivalent to the original $\phi_1$ and $\phi_2$ and a few moments consideration of (\ref{Chong1}) should convince the reader that as long as the latter both have periods of $2 \pi$ we will have no conical singularities away from the bubble.  However, at the bubble itself, the directions $\partial/{\partial \phi_1}$  and $\partial/{\partial \phi_2}$ degenerate (since $\partial/{\partial \psi}$ does) and we must use a coordinate which makes sense there, namely $\phi$.  Near $\theta = 0$ the metric on the bubble becomes
\be
ds^2 = \frac{R_0 + a^2}{\Xi_a} [ d \theta^2 + K_4 \, \theta^2 \, (d \phi+K_5 dt)^2 +K_6 dt^2] + \ldots
\ee
where
\be
K_4 =  \frac{(m_0 \delta - m_1 \beta)^2}{m_1^2}
\ee
and we will not be concerned with the precise value of the constants $K_5$ and $K_6$.  Recalling that $\phi$ is periodic at the bubble with period 
\be
\Delta \phi = \frac{2 \pi}{ \vert m_0 \delta - m_1 \beta \vert}
\ee
then we have a $Z_{\vert m_1 \vert}$ orbifold singularity at $\theta = 0$.  Likewise, one finds near $\theta = \pi/2$ one finds a $Z_{m_0}$ orbifold singularity on the bubble.   Hence, aside from the case $\vert m_1\vert = m_0 = 1$, when, as we will show below, the minimal $S_2$ is round, we have an orbifold singularity on at least one pole of the $S_2$.  One might describe this shape generically as an asymmetric football.  While classically this is a lack of smoothness, in string theory these singularities do not concern us, at least if one does not allow an additional orbifold singularity along the entire surface of the bubble (i.e. take $k = 1$).   The more generic case involves orbifold singularities within orbifold singularities and we do not know that this situation has ever been studied carefully.

\subsection{Causal stability}

We now check that these solutions have no closed timelike curves.  One simple way to rule out such pathologies is to show that $g^{t t}$ is everywhere negative definite, for then $t$ is globally a good time function.  One can check that $g^{t t}$ is independent of the choice of $\alpha$, $\beta$, $\gamma$, and $\delta$ and so we obtain the same manifestly negative definite expression obtained in \cite{RossBubbles}
$$
g^{t t} = -\frac{(R-R_0)}{W \Delta_\theta \rho^2 a^2 b^2} \Bigg[ a^2 b^2 \Xi_a \Xi_b (R-R_0) \Big[ R-R_0+\rho_0^2 +\sqrt{(a^2-b^2)^2+4 a^2 b^2 Q} \Big]
$$
\be \label{inversegtt}
+ \delta_0 \Big[ \Xi_b a^2 (b^2+R_0) \sin^2 \theta + \Xi_a b^2 (a^2+R_0) \cos^2 \theta \Big] \Bigg]
\ee
where $\rho_0^2 = \rho^2(R = R_0)$,
\be
\delta_0 = \frac{1}{2} \Big[ (a^2-b^2)^2 + 2 a^2 b^2 Q + (a^2 \Xi_b + b^2 \Xi_a) \sqrt{(a^2-b^2)^2 + 4 a^2 b^2 Q} \Big]
\ee
and we recall $R_0 + a^2 > 0$ and $R_0 + b^2 > 0$ (a necessary condition for $\rho^2 > 0$, as well as manifest from the expression for $R_0$ (\ref{bubblerad})).  The reader might be concerned the expression (\ref{inversegtt}) might vanish at $R = R_0$ but recall that $W(R_0) = 0$ and $R = R_0$ is the largest zero of $W$, so $g^{t t}$ is negative definite.

\subsection{Smoothness of the gauge potential and local charge}

If the gauge potential is to be finite at the bubble then it must be true that
\be
A_{\psi} = A \cdot \Big(\frac{\partial}{\partial \psi} \Big)
\ee
vanishes at the bubble since the Killing vector $\partial/{\partial \psi}$ does.   Note if one does not do this there will be a $\delta$-function flux along the bubble, since if one considers a disk in the $(R, \psi)$ plane near the bubble
\be
 \int_{\mathrm{disk}} F = \int_{\mathrm{boundary}} A
\ee
does not become small for $R$ arbitrarily close to $R_0$.   In a non-gravitational theory one often allows such nonzero $\delta$-function fluxes, provided they are suitably quantized in units of the electric charge of fundamental fermions.  This is simply the usual Dirac string construction, although in this case one has a sphere rather than a string.  However, once gravitational backreaction is included, the metric will not be smooth for such fluxes.  We wish to find entirely smooth solutions and so it must be true for the entire surface of the bubble that
\be \label{C0condit}
C_0 =  -\frac{2 \sqrt{3} s_{\epsilon}  a^2 b Q  \alpha}{\Xi_a [ b^2 - a^2 + \sqrt{(b^2-a^2)^2 + 4 a^2 b^2 Q} ] }
\ee
Note $C_0$ is nonzero unless $Q$, $a$, or $b$ vanishes and, as noted above, any of these options only lead to singular solutions.  

Likewise there are two other axes where we must ensure the potential does not have a hidden singularity.  The directions $\psi$ and $\phi$ become degenerate at $\theta = 0$ and $\theta =  \pi/2$.  This simply reflects the fact the axis $\phi_1$ degenerates at $\theta = 0$ and the axis $\phi_2$ at $\theta = \pi/2$.  As before, the places where these axis run into the surface of the bubble are simply the northern and southern poles of the minimal $S_2$.  Then we must ensure that
\be \label{Aaxis1}
 0 = A \cdot  \frac{\partial}{\partial \phi_1} \Big \vert_{\theta = 0} \rightarrow \delta C_0 - \gamma C_1 \vert_{\theta = 0} = 0
\ee
and
\be \label{Aaxis2}
 0 = A \cdot  \frac{\partial}{\partial \phi_2} \Big \vert_{\theta = \pi/2} \rightarrow -\beta C_0 + \alpha C_1 \vert_{\theta = \pi/2} = 0
\ee
If $C_0$ and $C_1$ were constant (and not identically zero) over the entire spacetime this would imply $\alpha \delta - \beta \gamma = 0$ and we would not have had a valid diffeomorphism in the first place.   The solution is that the gauge potential cannot be defined globally but must be defined in patches around $\theta = 0$ and $\theta = \pi/2$.  Perhaps the simplest such patches are two hemispherical regions between $0 \leq \theta < \theta_0$ and $\theta_0 < \theta \leq \pi/2$ respectively, for some constant $\theta_0$ between 0 and $\pi/2$.   Note these patches are not localized the radial direction; they extend throughout the entire spacetime.  In analogy to the terminology familiar from the magnetic monopole in four dimensions, one may refer to the patch surrounding $\theta = 0$ as the northern patch and around $\theta = \pi/2$ as the southern patch.   Since, as noted above, keeping the potential regular at the bubble forces $C_0$ to have the same value in both these patches we will be forced to take $C_1$ to have different values in these two different patches.  In particular if we wish a smooth solution we must take
\be \label{C1Ncondit}
C^{(N)}_1 = \frac{\delta}{\gamma} C_0 = - \frac{2 \sqrt{3} \, s_{\epsilon} a b^2 Q \,  \delta }{\Xi_b [ a^2 - b^2 + \sqrt{(b^2-a^2)^2 + 4 a^2 b^2 Q} ] }
\ee
and
\be \label{C1Scondit}
C^{(S)}_1 = \frac{\beta}{\alpha} C_0 = - \frac{2 \sqrt{3} \, s_{\epsilon} a^2 b Q  \, \beta}{\Xi_a [ b^2 - a^2 + \sqrt{(b^2-a^2)^2 + 4 a^2 b^2 Q} ] }
\ee
Alternatively, if one did not mind a $\delta$-function field flux along the $\theta = 0$ or $\theta = \pi/2$ axis one could define $C_1$ globally--this is simply the familiar Dirac string.  We prefer entirely smooth solutions so instead choose to work with these two patches.    To make contact with our original constants for the potential $K_0$ and $K_1$ (\ref{potorig}), the above conditions imply that at $\theta = 0$, $K_0 = 0$ but $K_1 \neq 0$ and likewise at $\theta = \pi/2$, $K_1 = 0$ but $K_0 \neq 0$.

One might not necessarily trust the above to ensure that the potential is entirely regular on the bubble at the poles of the $S_2$ since, as noted before, the Killing vectors in (\ref{Aaxis1}) and (\ref{Aaxis2}) are degenerating there.  Then let us note that along the axis $\theta = 0$
\be \label{Aphi1}
A_{\phi} (\theta = 0) = A \cdot  \frac{\partial}{\partial \phi} \Big \vert_{\theta = 0} = \frac{\sqrt{3} a b^2  s_{\epsilon} Q \delta}{\Xi_b (R + a^2)} + C_1^{(N)}
\ee
and along the axis $\theta = \pi/2$
\be \label{Aphi2}
A_{\phi} (\theta = \pi/2) = A \cdot  \frac{\partial}{\partial \phi} \Big \vert_{\theta = \pi/2} = \frac{\sqrt{3} a^2 b  s_{\epsilon} Q \beta}{\Xi_a (R + b^2)} + C_1^{(S)}
\ee
Provided that one specifies $C_1^{(N)}$ and $C_1^{(S)}$ as in (\ref{C1Ncondit}) and (\ref{C1Scondit}), it is straightforward to see using (\ref{bubblerad}) that at $R = R_0$,  $A_{\phi}$ vanishes at $\theta = 0$ and at $\theta = \pi/2$ as it should.  Since the bubble and the $\theta = 0$ and $\theta = \pi/2$ axis are the only places where $\psi$ and $\phi$ (and equivalently $\phi_1$ and $\phi_2$) degenerate and the potential is manifestly regular elsewhere, the conditions (\ref{C0condit}), (\ref{C1Ncondit}), and (\ref{C1Scondit}) are the necessary and sufficient conditions to have a regular gauge field.   In particular, they are sufficient to ensure $A_a A^a$ is regular everywhere; the conditions we have actually used are rather stronger than the criterion that $A^2$ is regular since conceivably one could have unexpected cancellations in the latter quantity.

The fact that the potential cannot be globally well defined follows from the fact that as soon as there is a minimal two-surface $\mathcal{S}$ (i.e. a bubble) one can define a nonzero charge by integrating the two form field strength over that surface
\be
q_m = \frac{1}{4 \pi} \int_{\mathcal{S}} F
\ee
This charge will be conserved since $F$ is closed, at least as long as the spacetime does not evolve in such a way that $\mathcal{S}$ ceases to exist.  Note further that one will obtain the same charge from any other two-surface cobordant to $\mathcal{S}$, again since $F$ is closed.  This may be described as a local charge since in five dimensions the globally conserved charges are defined by integrating forms over a three manifold that is topologically $S_3$, usually just the $S_3$ at infinity.  A global electric charge is given by integrating the dual of $F$ (plus, in the context of minimal supergravity, a Chern-Simons term) and a global magnetic charge by integrating a three-form.   While, as we discuss later, one may obtain an electric charge for these solutions, the global magnetic charge vanishes identically since there is no three-form field strength.  This local charge cannot (apparently) be given in terms of such a three-form and in any case the gauge symmetries seem to be entirely accounted for by the usual global charges.  In the context of black rings, such charges were dubbed ``dipole charges'' \cite{EmparanDipBR04}, where $\mathcal{S}$ is (or cobordant to) the $S_2$ of the horizon ($S_1 \times S_2$),  although the name derives from the fact that in that context one obtains $q_m$ with opposite signs from points on opposite sides of the ring.  Also to measure this charge for black rings the surface $\mathcal{S}$ must go through the middle of the ring; for our charge, in contrast, one still gets a nonzero answer with an $S_2$ everywhere arbitrarily far away from the bubble.

For the solitons we are considering, taking $\mathcal{S}$ to be the bubble surface one finds 
\be
q_m = - \frac{\sqrt{3} (b^2+R_0)}{2 a m_0 \Xi_b}
\ee
where the orientation of the spacetime is, as before, $\epsilon_{t \phi_1 \phi_2 \theta r} = s_{\epsilon} \sqrt{-g}$. In the case $b^2 = a^2$ the local charge is directly related to $Q$ and, as we will later see, the electric charge
\be
q_m = -  \frac{ \sqrt{3 Q} \, a}{2 m_0 \Xi_a}
\ee
while in the more generic case the relation of $q_m$ and $Q$ is somewhat more indirect
\be
q_m =  - \frac{\sqrt{3}[ b^2 - a^2 + \sqrt{(b^2-a^2)^2+4 a^2 b^2 Q}]}{4 a m_0 \Xi_b}
\ee
Note the local charge $q_m$ is necessarily nonzero for any smooth solutions -- just as we found above, none of the solutions have a globally defined regular potential.  This connection, of course, is no accident.  If there were a globally well-defined potential then the integral of $F$ over any compact manifold would necessarily vanish.  In fact, on the surface of the bubble we locally have a situation identical to the magnetic monopole in four dimensions and, as in that case, one necessarily finds either one must take the potential to be defined in patches or the potential and the field strength diverges along some axis.  

Recall fundamental fermions with charge $e_0$ pick up a phase $e^{i e_0  \oint A \cdot dl}$ when moved along a closed loop.  We must ensure such fermionic wavefunctions are continuous between our two patches and hence that they pick up the same phase when going around a closed loop.  Alternatively, one can allow the patches to overlap in some region and demand that the acquired phase be the same no matter which potential one considers in the overlap region.  In any case, one finds
\be \label{Diraccircle}
 2 \pi n   = \Delta \phi (C^{(N)}_1 - C^{(S)}_1) e_0 = s_{\epsilon} \mathrm{sgn}[ \alpha (\alpha \delta - \beta \gamma)] 4 \pi q_m e_0 
\ee
for some integer $n$.  Absorbing the various signs into the definition of $n$ yields
\be
q_m e_0 = \frac{n}{2}
\ee
the usual Dirac quantization condition.  That one would obtain the condition familiar from four dimensions (with $q_m$ instead of the global magnetic charge) should be no surprise since, as noted above, on the surface of the bubble the situations are identical.  For the sake of compactness, it is handy to define another integer $p = -n$.  We then obtain
\be \label{Dirac0}
p = \frac{\sqrt{3} e_0 [ b^2 - a^2 + \sqrt{(b^2-a^2)^2+4 a^2 b^2 Q}]}{2 a m_0 \Xi_b} = \frac{\sqrt{3} e_0 (b^2 + R_0)}{a m_0 \Xi_b}
\ee

Depending on the context, there may be some special value of $e_0$ which one wishes to impose.  For example, one may wish to embed the solution into $AdS_5 \times S_5$ via the method of \cite{CveticEmbed} where the gauge field $A$ is just a Kaluza-Klein gauge field
\be
ds^2_{10} = ds^2_5 + \frac{1}{g^2} {\Sigma}_{i} [ d \mu_i^2 + \mu_i^2 (d \phi_i + \frac{g}{\sqrt{3}} A)^2 ]
\ee
and the coordinates on the $S_5$ are $\mu_i$ and $\phi_i$.  In this context, one is forced to take
\be
 e_0 = \frac{g}{\sqrt{3}}
 \ee
  For our asymptotically AdS solitons we will later impose this condition.  We wish to emphasize, however, that once one requires $e_0 = g/\sqrt{3}$, or more generically any condition where $e_0$ is proportional to $g$, the limit $g \rightarrow 0$ of (\ref{Dirac0}) will no longer be a sensible condition to impose.  In such a limit one obtains only neutral fundamental fermions, in which case one never had a Dirac quantization condition to begin with.  As we will see later, the asymptotically flat solutions often have qualitatively different features from the asymptotically AdS solitons.

Let us contrast the above quantization conditions with those one would obtain if one refused to admit the constants $C_0$ and $C_1$.  Let us define the corresponding potential as $\bar{A}$
\be
\bar{A}  = s_{\epsilon} \frac{\sqrt{3} q}{\rho^2} \Big( \frac{\Delta_\theta }{\Xi_a \Xi_b} dt - \omega \Big) = A \Big \vert_{C_0 = C_1 = 0}
\ee
One then cannot set $\bar{A}_{\psi} (R_0) = 0$ but if one takes the perspective, as in \cite{RossBubbles}, that this is not problematic provided the flux is suitably quantized in terms of the charge of fundamental fermions, then one finds a Dirac quantization condition analogous to (\ref{Diraccircle}):
\be
2 \pi \bar{n} = \Delta \psi \bar{A}_{\psi} (R_0) e_0
\ee
and defining $\bar{p} = \mathrm{sgn}[ \alpha ] s_{\epsilon} \bar{n}$
\be
\bar{p} = \frac{\sqrt{3} e_0 m_0 a^2 b Q}{\Xi_a (b^2 + R_0)} = \frac{\sqrt{3} e_0 m_0 (a^2 + R_0)}{b \Xi_a}
\ee
after a few lines of algebra using the value of $R_0$ from eq. (\ref{bubblerad}).  Comparing this to what one obtains with our quantization condition (\ref{Dirac0}), one finds the ratio between the two is proportional  to $\gamma/\alpha = m_1/m_0$  (\ref{condit3})
\be
\bar{p} = m_0^2 \frac{\gamma}{\alpha} p = m_1 m_0 p
\ee
Hence the quantization conditions are equivalent only if $\vert m_1 \vert = m_0 = 1$.  As we will discuss later, this turns out to be the condition if and only if the bubbles have equal magnitude angular momenta in the two orthogonal spatial planes.  In any other case, generic values of $\bar{p}$ will not correspond to integer values of $p$ and our condition will be violated.   From the perspective of allowing these $\delta$-function fluxes, the stronger condition we impose (\ref{Dirac0}) comes from a second Dirac quantization condition arising from the fact the potential is not globally defined due to $q_m \neq 0$.  One can also see this from (\ref{Aphi1}) and (\ref{Aphi2}); if $C_1 = 0$ the potential on the poles of the $S_2$ is unequal and at the surface of the bubble the argument proceeds entirely along the lines familiar from the magnetic monopole in four dimensions.

\subsection{Summary}

It is worth collecting the criteria we must impose to obtain smooth solutions at this point.  We began with a four continuous parameter family of black hole solutions $(m, q, a, b)$ (such that $a^2 g^2 < 1$ and $b^2 g^2 < 1$) and added an additional parameter $\gamma/\alpha$ by different choices of cycles one can pinch off.   We also allowed some additional constants $(C_0, C_1)$ in the gauge potential, but found regularity fixed them entirely.  It will be handy to define
\be
\gamma_0 \equiv \frac{\gamma}{\alpha} = \frac{m_1}{m_0}
\ee
Then from (\ref{condit3}) to pinch off a cycle and produce a soliton we must require that
\be \label{SC1}
\gamma_0 = \frac{a(1 - b^2 g^2) [a^2 - b^2 + \sqrt{(a^2-b^2)^2+4 a^2 b^2 Q}]}{b (1 - a^2 g^2) [b^2 - a^2 + \sqrt{(a^2-b^2)^2+4 a^2 b^2 Q]}}
\ee
where recall the dimensionless charge Q is defined so that $q = -a b Q$.  The bubble is located at $r^2 = R = R_0$ where
\be
R_0 = \frac{1}{2} \Big[ - (a^2+b^2) +  \sqrt{ (a^2-b^2)^2 + 4 a^2 b^2 Q} \Big].
\ee
and the value of $m$ is  (\ref{condit4})
\be \label{summmass}
m = \frac{Q}{4} \Big[ a^2 + b^2 + 2 a^2 b^2 g^2 + \sqrt{ (a^2 - b^2)^2 + 4 a^2 b^2 Q} \Big].
\ee
 We will not have any curvature singularities (due to encountering a zero of $\rho$) provided
\be \label{summQ}
0 < Q < 1
\ee
From the absence of a conical singularity on the entire bubble surface (\ref{conic1})
\be \label{SC2}
-\frac{ R_0 \Big[ (a^2-b^2)^2 + 2 a^2 b^2 Q + (a^2 \Xi_b + b^2 \Xi_a) \sqrt{(a^2-b^2)^2 + 4 a^2 b^2 Q} \Big]^2}{a^2 b^4 \Xi_a^2 \Big[b^2 - a^2 + \sqrt{(a^2-b^2)^2 + 4 a^2 b^2 Q} \Big]^2} = \frac{1}{m_0^2 k^2}
\ee
and from the Dirac quantization condition (\ref{Dirac0})
\be \label{SC3}
\frac{ \sqrt{3} (b^2+R_0) e_0 }{a m_0 \Xi_b} = p
\ee
Solving for these conditions (\ref{SC1}), (\ref{SC2}), and (\ref{SC3}) then fixes the remaining original continuous parameters of the black hole $(q, a, b)$ in terms of four integers $(m_0, m_1, p, k)$ and the fermion charge $e_0$.  

\setcounter{equation}{0}
\section{Asymptotic charges}

\subsection{Gravitational Charges}

For the black hole solutions we began with, it is possible to derive the mass as a result of the first law, as is done in \cite{CCLP05}.  For these solitons with local charge, one might worry that such a first law should include a term due to local charge, as occurs for dipole black rings \cite{CopseyHorowitz}, and integrating the full first law might lead to results different from those of \cite{CCLP05}.  Further, without a horizon the appropriate definition of $\Omega$ becomes somewhat confusing.   To avoid these complications, we will present simple and efficient geometric definitions of the conserved charges, namely the Komar integrals for the asymptotically flat spaces and the electric part of the Weyl tensor in the AdS case that we trust will be regarded as entirely noncontroversial.  We will also provide results for the charges based on covariant phase space Lagrangian methods.   In the end, the above concerns appear to be unfounded, at least for this class of solutions; all these methods reproduce precisely the results of \cite{CCLP05}.

Let us first consider charges in the asymptotically flat case. It is straightforward to check that the 2-form $F$ falls off fast enough that the Komar integrals corresponding to a Killing vector are conserved.   
With the normalization conventions\footnote{We associate the angular momenta $J_{\phi_i}$ to minus the generator $\frac{\partial}{\partial \phi_i}$ which differ by a sign from the convention used in \cite{MP}.} of  \cite{Iyer:1994ys, GMT99}
\be
M = -\frac{3}{16 \pi G} \int \star \nabla \xi_0 = -\frac{3}{32 \pi G}  \int \epsilon_{ a b c d e} \nabla^d \xi^e =  \frac{3 \pi m}{4 G}
\ee
where $\xi_0 = \partial/{\partial t}$.  The lack of curvature singularities  (\ref{summQ}) forces $ m > 0$ (\ref{summmass}). Hence, we have manifestly positive mass solitons.  

For the angular momenta associated with the Killing vectors $\xi_1 = -\frac{\partial}{\partial \phi_1}$ and $\xi_2 = -\frac{\partial}{\partial \phi_2}$, respectively, 
we find
\be
J_{\phi_1} =  -\frac{1}{8 \pi G} \int \star \nabla \xi_1 =  \frac{\pi a \, (2  m - b^2 Q) }{4 G}
\ee
and
\be
J_{\phi_2} =  -\frac{1}{8 \pi G} \int \star \nabla \xi_2 = \frac{\pi b \, (2 m -a^2 Q) }{4 G}
\ee
Inputting the value of $m$ required from (\ref{condit4}) for solitons gives
\be
J_{\phi_1} =  \frac{ \pi a Q}{8 G} \Big[ a^2 - b^2 +\sqrt{(a^2-b^2)^2 + 4 a^2 b^2 Q} \Big]
\ee
and
\be
J_{\phi_2} = \frac{ \pi b Q}{8 G} \Big[ b^2 - a^2 +\sqrt{(a^2-b^2)^2 + 4 a^2 b^2 Q} \Big]
\ee
and hence these solutions are necessarily rotating in both planes.  

It is known that nearly all ways to define conserved quantities in AdS space are  equivalent  up to zero-point ambiguities \cite{HollandsIshibashiMarolf}.  Perhaps the computationally easiest way to find charges for $AdS_d$ is via the electric part of the Weyl tensor \cite{AshtekarDas}
\be \label{weyl1}
Q_{\xi} = \frac{- l^3}{8 \pi G (d -3)} \int dS E_{ a b} u^a \xi^b
\ee
where the integral is over the boundary at spatial infinity of a spacelike slice $\Sigma$, $dS$ contains the usual measure on that boundary (i.e. in global coordinates grows as $r^{d -2}$) and $u^a$ is the timelike unit normal to $\Sigma$.  The electric part of the Weyl tensor is
\be \label{weyl2}
E_{ a b} = C_{a c b d} \frac{\nabla^c \Omega}{\Omega} \frac{\nabla^d \Omega}{\Omega}
\ee
and the Weyl tensor is, as usual,
\be \label{weyl3}
C_{a c b d} = R_{a c b d} - \frac{g_{a b} R_{c d} - g_{a d} R_{b c} - g_{b c} R_{a d} + g_{c d} R_{ a b}}{d-2} + \frac{ R (g_{a b} g_{c d} - g_{a d} g_{b c})}{(d-1) (d-2)}
\ee
The factor $\Omega$ is a conformal completion for the AdS space, so that the unphysical metric
\be
\tilde{g}_{a b} = \Omega^2 g_{ a b}
\ee
given in terms of the physical metric $g_{ a b}$ admits a smooth limit at infinity.  In global coordinates, one typically takes
\be
\Omega = \frac{1}{r}
\ee
although of course there is an infinite family of other completions one might use.  Note in the above all quantities in (\ref{weyl1}-\ref{weyl3}) are given, for ease of computational use, in terms of the physical metric.  One can quickly check that $F_{a b}$ again falls off fast enough \cite{AshtekarDas} such that we obtain conserved charges. Using the electric part of the Weyl tensor, for the mass one finds
\be
M = Q_{t} = \frac{ \pi  [ (2 \Xi_a + 2 \Xi_b - \Xi_a \Xi_b) m - 2 a^2 b^2 Q g^2 (\Xi_a + \Xi_b)]}{4 G \, \Xi_a^2 \Xi_b^2}
\ee
Note for our solitons, while $m > 0$ since $Q > 0 $ the mass might appear to be in danger of becoming negative.  A bit of algebra, however, shows that as long as one enforces the absence of curvature singularities (\ref{condit5}, \ref{condit8}), $M$ will be positive definite.
For the angular momenta\footnote{Associating the angular momenta with minus the generator $\frac{\partial}{\partial \phi_i}$; see, e.g. \cite{GutowskiReall}, for a discussion of this point.}
\be
J_1 = Q_{\phi_1} =  \frac{\pi a \, ( 2 m - b^2 Q (1+ a^2 g^2))}{4 G \, \Xi_a^2 \Xi_b}
\ee
and
\be
J_2 =  Q_{\phi_2} =  \frac{ \pi  b \, ( 2 m - a^2 Q (1+ b^2 g^2))}{4 G \, \Xi_b^2 \Xi_a}
\ee
 For our solitons, inputting the value of $m$, this becomes 
\be
J_{\phi_1} =  \frac{ \pi a Q}{8 G \Xi_a^2 \Xi_b } \Big[ a^2 - b^2 +\sqrt{(a^2-b^2)^2 + 4 a^2 b^2 Q} \Big]
\ee
and
\be
J_{\phi_2} = \frac{\pi b Q}{8 G \Xi_b^2 \Xi_a } \Big[ b^2 - a^2 +\sqrt{(a^2-b^2)^2 + 4 a^2 b^2 Q} \Big]
\ee
and, just as in the asymptotically flat case, any smooth solitons are necessarily rotating in both planes.  Note the asymptotically flat charges are precisely reproduced as  $g \rightarrow 0$.   Further, comparing the angular momenta between the two planes
\be \label{Jdiff}
J_{\phi_1} \mp J_{\phi_2} =  \frac{\pi (a \mp b) Q}{8 G \Xi_a^2 \Xi_b^2} \Big[ (a \pm b)^2 (1 \mp a b g^2) + (1 \pm a b g^2) \sqrt{(a^2-b^2)^2+4 a^2 b^2 Q} \Big] 
\ee
Since the term in brackets in (\ref{Jdiff}) is positive definite (since $\vert a g \vert < 1$ and $\vert b g \vert < 1$),  $J_{\phi_1} = \pm J_{\phi_2}$ if and only if $b = \pm a$.  For the sake of compactness, we will use the term ``equally rotating'' to describe solutions with equal magnitudes of angular momenta in the two planes (i.e.  $\vert J_{\phi_1} \vert = \vert J_{\phi_2} \vert$) since any distinctions between the two signs should be clear in the given context.

While the above methods have the advantage of being calculationally and conceptually rather straightforward, they can be shown to be canonically associated with the Killing vectors in the Hamiltonian or Lagrangian sense only indirectly, see e.g. \cite{Iyer:1994ys,HollandsIshibashiMarolf}. One can directly calculate the covariant phase space charges for both asymptotically flat and asymptotically AdS spacetimes in a unified framework via the methods of \cite{Barnichetal1}.  In this context, the zero-point ambiguities are fixed by setting the charges to zero for asymptotically flat and globally asymptotically AdS space.  The method consists of the integration of a $3$ form associated with a Killing vector $\xi$ defined uniquely from the theory at hand independently of the asymptotic behavior of the spacetime. The charge is defined by the integration of a form and so is coordinate-independent. The charge could be non-zero only when the surface of integration has a non-trivial homology, i.e. surrounds a black hole, a conical deficit or a non-trivial topology of the spacetime for example. The exact expression of the surface charge for our Lagrangian can be found in \cite{Barnich:2005kq}. 

Since the gauge field decays for large radius, it is convenient to evaluate the mass and angular momenta on a sphere at large radius so that only the gravitational field will contribute. Since the metric is a smooth function of $g$ in any region of the spacetime around some fixed radius, the surface charge constructed out of the metric will also depend smoothly on $g$ in that region and the limit $g \rightarrow 0$ will be well defined. By construction, the charges associated with a Killing vector $\xi$ will be the canonical quantities associated with $\xi$ in the limit $g \rightarrow 0$ as well. Using a Mathematica code, one finds the same charges as above.
 
 \subsection{Electric charge and the BPS bound}

The usual definition of electric charge, consisting of an integral over the $S_3$ at infinity, in the asymptotically AdS case yields
\be \label{usualQE}
Q_{E} = \frac{1}{16 \pi G } \int_{S_\infty^3} (\star F - F \wedge A/\sqrt{3} ) = - \frac{\sqrt{3} \pi s_{\epsilon} a  b Q}{4 G   \, \Xi_a \Xi_b}
\ee
and, as one might expect, defining charge in the same way in the asymptotically flat case by
\be
Q_{E}  = - \frac{ \sqrt{3} \pi s_{\epsilon} a b Q}{4 G}
\ee
Note in this context one must define the potential (and integral) in patches but since the field strength asymptotically goes to zero the Chern-Simons term  drops out of the expression.  Such an expression would not yield a sensible conserved charge at any finite radius. More precisely, one will not obtain the same asymptotically conserved charge. Since the potential is not globally well-defined one picks up an extra contribution from the interface between the two patches and what one would like to call the electric charge depends on the choice of surface.    However, as long as the above expressions are evaluated at infinity with the usual asymptotics (either flat or AdS) where the field strength falls off at infinity, it will be conserved and gauge invariant (under smooth gauge transformations continuously connected to the identity).  

It is worth noting that the reason one obtains nonzero electric charge for a completely regular solution without any sources or internal boundaries is entirely due to the fact that the potential is not globally well-defined.  Without this local charge, (\ref{usualQE}) would yield the conserved charge at any radius and provided the field strength and potential were regular everywhere the expression would vanish as the integration surface $\mathcal{S}$ approached the bubble and the three-volume vanished.  One can show the $Q_E$ above may be directly related to the difference in gauge potential between the two patches (Appendix III).

 It is possible to define another notion of charge that may be calculated at a finite radius in the presence of local magnetic charge, that is
\be
\bar{Q}_{E} = \frac{1}{16 \pi G } \int_{\mathrm{bdy} \, \, \, \mathrm{patches}} (\star F - F \wedge A/\sqrt{3} ) 
\ee
where the integral runs over the boundaries of all gauge patches needed to define the potential.  In our context this means one has contributions not only from an $S_3$, but also from the interface between the two gauge patch hemispheres all the way down to the bubble.  For solutions like ours, where the field equation is satisfied without additional sources or singularities, $\bar{Q}_E$ is necessarily zero.   More generically, in the presence of point charges or electrically charged black holes, for example, it will be nonzero.  While this definition of charge will be conserved in the presence of nonzero $F$ and a potential which is not globally well-defined, it is not the usual notion which, among other things, enters into the conventional BPS bounds.   It also does not match the usual notion of a conserved charge in that it is not calculated only in the asymptotic region.  Hence it is not clear that this notion is anything more than a curiosity.  For the remainder of our discussion we will return to the conventional definition (\ref{usualQE}).

For the asymptotically flat case, given the normalizations above for our conserved quantities and in the action (\ref{Sact}), the BPS bound \cite{GKLTT, GMT99} is \footnote{In both the asymptotically flat and asymptotically AdS cases we will use the conventional definition of the BPS bound; various possible modifications are left for the discussion at the end of the paper.}
\be \label{BPSasymflat}
M \geq \sqrt{3} \vert Q_E \vert
\ee
One fairly straightforward way to verify that, including all normalizations, (\ref{BPSasymflat}) is precisely correct is to compare it with a known supersymmetric solution.  Of course, such solutions  saturate (\ref{BPSasymflat}).  In five asymptotically flat dimensions, probably the simplest such solution is the extremal static charged black hole \cite{GKLTT}--that is the generalization of Reissner-Nordstrom to five dimensions.

In terms of the parameters of the present solution the BPS bound (\ref{BPSasymflat}) in the case $g = 0$ is equivalent to the statement that
\be \label{flatBPS1}
\beta_Q \equiv  \frac{\sqrt{3} \vert Q_E \vert}{M} =  \frac{\vert q \vert}{m} \leq 1
\ee
or for these solitons if one defines $b_0 = b/a$
\be
1 \leq \frac{ 1 + b_0^2 + \sqrt{(1-b_0^2)^2+4 b_0^2 Q}}{4 \vert b_0 \vert}
\ee
Equivalently the bound will be violated if 
\be \label{flatBPS2}
\frac{5 - Q - \sqrt{(9-Q) (1-Q)}}{4} < \vert b_0 \vert < \frac{5 - Q + \sqrt{(9-Q) (1-Q)}}{4} 
\ee
saturated if  
\be
\vert b_0 \vert = \frac{5 - Q \pm  \sqrt{(9-Q) (1-Q)}}{4} 
\ee
and otherwise respected.  Noting that the lower bound in (\ref{flatBPS2}) is strictly less than one and the right hand side strictly greater than one, there is always some range around the equally rotating case which violates the bound.  In particular, the bound is violated in the equally rotating case.

In the AdS case, the BPS bound is given by (\cite{AdSBPS}, \cite{GutowskiReall})
\be
M \geq \sqrt{3} \vert Q_E \vert + g \vert J_{\phi_1} \vert + g \vert J_{\phi_2} \vert\ee
 or equivalently the statement that
 \be
 \beta_Q \equiv \frac{\sqrt{3} \vert Q_E \vert + g \vert J_{1} \vert + g \vert J_2 \vert}{M} \leq 1
 \ee
One may check the normalizations in the above with the relatively simple supersymmetric black holes of \cite{GutowskiReall}.  For our solitons\footnote{Note the rather simpler expressions, in particular for BPS saturating solutions, derived in \cite{CCLP05} depend crucially on the signs of electric charge and angular momenta all being positive and for our solitons this is often not true.}
 \be
 \beta_Q = \frac{12 \vert a b \vert \, \Xi_a \Xi_b + 2 (a^2 - b^2)( \vert a g \vert \, \Xi_b - \vert b g \vert \, \Xi_a) + 2 x_0 ( \vert a g \vert \, \Xi_b + \vert b g \vert \, \Xi_a) }{(2 \Xi_a + 2 \Xi_b - \Xi_a \Xi_b)(a^2+b^2 + 2 a^2 b^2 g^2+x_0) - 8 a^2 b^2 g^2 (\Xi_a + \Xi_b)}
 \ee
where
\be
x_0 = \sqrt{(a^2-b^2)^2+4 a^2 b^2 Q}
\ee

It is difficult to make any analytic statements analogous to the asymptotically flat case, although plotting several examples the pattern seems rather similar; there are regions which exceed the BPS bound if the angular momenta have nearly equal magnitude and other regions where the bound is satisfied.   Whether the BPS bound is in fact violated for smooth solutions depends on whether one considers equally rotating, unequally rotating asymptotically flat, or unequally rotating asymptotically AdS solutions and we now turn to a case by case analysis of these solitons.

\setcounter{equation}{0}
\section{Equally Rotating Bubbles}

\subsection{The solution}

From (\ref{SC1}) if $b = \pm a$, independently of the value of $Q$
\be
\gamma_0 =\frac{m_1}{m_0} = \pm 1
\ee
and so
\be
\vert m_1 \vert = m_0 = 1
\ee
The solution in this case is simple enough that writing the metric and potential explicitly may be useful for the reader.  For the sake of simplicity, we define rescaled angles
\be
\bar{\theta} = 2 \theta
\ee
\be
\bar{\psi} = 2 \alpha \psi
\ee
and
\be
\bar{\phi} = - (\beta \mp \delta) \phi
\ee
so that $ 0 \leq \bar{\theta} \leq \pi$, $\bar{\psi}$ has a period of $4 \pi$, and $\bar{\phi}$ a range of $2 \pi$.  Note the factor we use to rescale $\phi$ is always nonzero, since if $\beta = \pm \delta$,  $\alpha \delta - \beta \gamma  = 0$.  In these rescaled coordinates the metric is
$$
ds^2 = g_{t t} (r) dt^2 + \alpha_1 (r) dt (d \bar{\psi} + \beta_0 d \bar{\phi}) + \alpha_2 (r)  (d \bar{\psi} + \beta_0 d \bar{\phi})^2
$$
\be \label{eqrotmet1}
+ \frac{ r^2 + a^2}{ 4 \Xi_a}(d\bar{\theta}^2 + \sin^2\bar{\theta} d \bar{\phi}^2) + \frac{r^2 (r^2+a^2)}{W} dr^2
\ee
while the potential is
\be
A = \pm \frac{ s_{\epsilon} \sqrt{3}  \, a^2 Q}{\Xi_a (r^2 + a^2)} \Big[-dt + \frac{a}{2}\Big(d \bar{\psi} + \beta_0  d \bar{\phi}\Big)\Big] + C_0 d\psi + C_1 d\phi
\ee
where
\be
\beta_0 =\cos(\bar{\theta}) - \frac{\beta \pm \delta}{\beta \mp \delta}
\ee
and
$$
g_{tt} (r) = -\frac{1}{\Xi_a^2 (r^2+a^2)^2} \Big[g^2 \Xi_a r^6 + \Xi_a (1 + 2 a^2 g^2) r^4
$$
\be \label{eqrotgtt}
+a^2 (2 - a^2 g^2 - a^4 g^4 -\Xi_a Q - Q^{3/2}) r^2+ a^4 (\Xi_a (1- Q) - Q^{3/2} + Q^2) \Big]
\ee
\be \label{alph1}
\alpha_1(r) = - \frac{ a^3 Q^{3/2} \Big[ r^2 + a^2 \Big(1 - \sqrt{Q}\Big)\Big]}{\Xi_a^2 (r^2+a^2)^2}
\ee
\be \label{alph2}
\alpha_2 (r) = \frac{\Big[ r^2+a^2 \Big(1 - \sqrt{Q}\Big)\Big]}{4 \Xi_a^2 (r^2+a^2)^2}  \Big[ \Xi_a r^4 + \Xi_a a^2 (2 + \sqrt{Q}) r^2 + a^4( \Xi_a (1 + \sqrt{Q}) + Q^{3/2})\Big]
\ee
and
\be
W = \Big(r^2+a^2 \Big)^2 \Big(1 + g^2 r^2\Big) - a^2 Q \Big( 1 + a^2 g^2 + \sqrt{Q}\Big) r^2 - a^4 Q \Big(2 - Q \Big)
\ee
Note the location of the bubble in this case is at
\be
R_0 = -a^2(1 - \sqrt{Q})
\ee
as one may see directly in (\ref{alph1}, \ref{alph2}).   In this case once $\bar{\psi}$  pinches off, one is left with a round $S_2$, as promised earlier.   Specifying $\beta = \mp \delta$ reproduces the usual Hopf fibration (Appendix I).  For the case $g = 0$, this solution can be generated by applying the solution generating technique of  (\cite{Mizoguchi:1998wv}-\cite{GSG2}, \cite{Clement:2008qx}-\cite{Gal'tsov:2008nz}) to the Eguchi-Hanson soliton \cite{EH}.

\subsection{Smoothness}

With $b = \pm a$ the absence of a conical singularity at the bubble (\ref{SC2}) becomes a cubic in $\sqrt{Q}$
\be \label{Eqcon}
\frac{\Big(1 - \sqrt{Q}\Big) \Big( 2 \Xi_a + \sqrt{Q}\Big)^2}{\Xi_a^2}= \frac{1}{k^2}
\ee
 If one defines
\be
\vartheta = \arccos \Big[ 1 - \frac{27 \, \Xi_a^2}{2 k^2 (1 + 2 \Xi_a)^3} \Big]
\ee
the single real solution of (\ref{Eqcon}) is
\be
Q_0 = \Bigg [\frac{2 (1 + 2 \, \Xi_a)}{3} \cos \Big(\frac{\vartheta}{3} \Big) +\frac{1 - 4 \, \Xi_a}{3} \Bigg]^2
\ee
One may further verify $ 0 < Q_0 < 1$ so we respect the bound imposed by (\ref{condit8}).

For the asymptotically flat case, the Dirac quantization condition is
\be
a = \frac{p}{e_0 \sqrt{3 Q}}
\ee
which merely quantizes the overall dimensionful scale and imposes no further restriction on the soliton.    For the asymptotically AdS case  the Dirac quantization condition becomes
\be \label{EqDirac3}
\frac{a g \sqrt{Q_0}}{1 - a^2 g^2} = p\frac{g}{\sqrt{3} e_0}
\ee
One can check that for any fixed value of $k$ the left hand side of (\ref{EqDirac3}) monotonically interpolates between 0 and $\pm \infty$ (depending on the sign of $a$), as $\vert a g \vert$ goes between 0 and 1, so for any given $p$ and $e_0$ there will be a unique value of $a g$ such that $0 < \vert a g \vert < 1$ and (\ref{EqDirac3}) is satisfied.  For example, if $e_0 = g/\sqrt{3}$ and $k = p =1$, $ \vert a g \vert \approx .6404$.  For the sake of visualization we have plotted the left hand side of (\ref{EqDirac3}) versus $a g$ for $k= 1$.  As $ \vert a g \vert \rightarrow 1$ $Q_0 \rightarrow 1$ and one gets to arbitrarily large values of p.
\begin{figure}[t]
\centering
	\includegraphics[scale= 1.6]{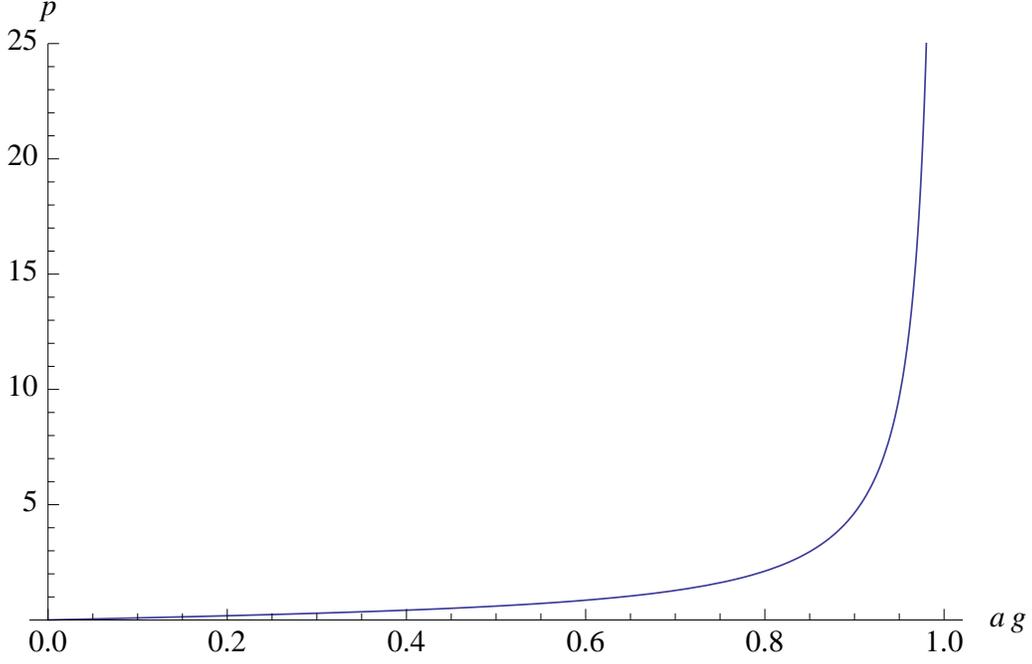}
	\caption{Dirac quantization condition for equally rotating $k = 1$, $e_0 = g/{\sqrt{3}}$; only integer values of p are actually realized.}
	\label{CaseIplot}
	\end{figure}
	
	\subsection{Ergoregion}
	
	We wish to find ergosurfaces associated to $\partial/{\partial t}$, or in other words the zeroes, if any, of $g_{t t}$.  Let us first note $g_{t t}$ (\ref{eqrotgtt}) is independent of the sign of $b/a$, so these questions are independent of the relative signs of the two angular momenta ($J_{\phi_1}$ and  $J_{\phi_2}$).  For the asymptotically flat case, the zeroes of $g_{t t} (r)$ occur at
	\be
	r^2 = \frac{a^2}{2} \Bigg[ -2 +Q + Q^{3/2}  \pm Q \sqrt{ (\sqrt{Q} - 1) (3 + \sqrt{Q})} \, \Bigg]
	\ee
	where the signs indicate the two possible roots (with any given sign of $b/a$).  Since regularity demands $0 < Q < 1$, both roots are complex and for  $g = 0$ there is no ergoregion. In this case, the Killing vector $\frac{\partial}{\partial t}$ is globally timelike.

	The asymptotically AdS case is somewhat more complicated.  If we define $r^2 = R_0 + a^2 z$ so that $z$ is a dimensionless measure of how far we are away from the bubble,
	\be
	g_{t t} = \frac{z_3 z^3 + z_2 z^2 + z_1 z + z_0}{\Xi_a^2 (z + \sqrt{Q})^2}
	\ee
	where
	\be
	z_3 = -a^2 g^2 \Xi_a
	\ee
	\be
	z_2 = - \Xi_a \Big[\Xi_a + 3 a^2 g^2 \sqrt{Q} \Big]
	\ee
	\be
	z_1 = \sqrt{Q} \Big[ - 2 \Xi_a^2 + \Xi_a (1- 3 a^2 g^2) \sqrt{Q} + Q \Big]
	\ee
	and
	\be
	z_0 = - \Xi_a^2 Q \Big[1 - \sqrt{Q} \Big]
	\ee
	Note $z_0$, $z_2$, and $z_3$ are negative definite.  Hence $g_{t t}$ is negative at the bubble surface, as well as at infinity.  Whether $z_1$ is positive or negative depends on the values of $Q$ and $a g$.  If $z_1$ is negative we are assured there is no ergoregion, while if it is positive there may be a bounded region where $g_{t t}$ becomes positive.  Note such ergoregions, provided they exist, are rather unusual; one has both an inner and outer ergosphere.  Aside from the usual black hole case where the inner ergosphere is inside a horizon and inaccessible to outside observers, we are not familiar with another solution with this structure of ergospheres.  Indeed, arguing that there is no ergoregion disjoint from the horizon is an important part of the black hole uniqueness theorems.

\begin{figure}[t]
	\begin{picture} (0,0)
    	\put(25,7){$g_{t t}$}
         \put(385, -118){$z$}
    \end{picture}

\centering
	\includegraphics[scale= 1.1]{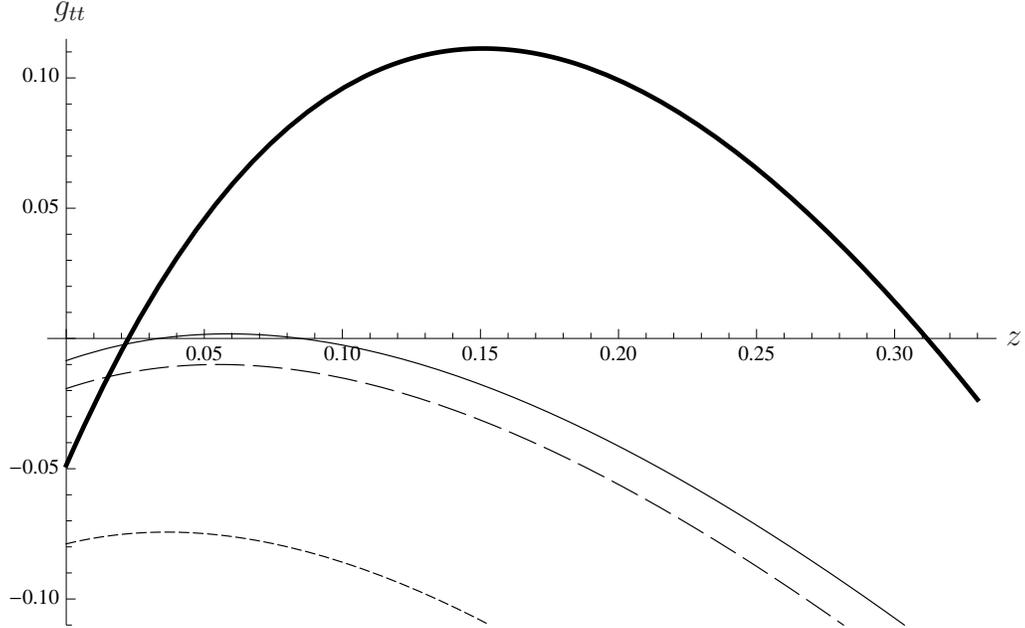}
	\caption{$g_{t t}$ for $(p = 1, k = 1)$ [short dashes], $(p = 1, k=2)$ [long dashes], $(p = 1, k=3)$ [thin line], and $(p = 2, k=1)$ [thick line] equally rotating AdS solutions with $e_0 =g/{\sqrt{3}}$; $z$ parametrizes the radial distance from the bubble.}
	\label{EqRoterg}
	\end{figure}

	In fact, such ergoregions do exist for certain values of the parameters.  Recall from the previous section that as $p$ becomes large for any fixed $k$, $\vert a g \vert \sim 1$.   Furthermore, as $\vert a g \vert \rightarrow 1$, $\Xi_a \rightarrow 0$ and $z_0$, $z_2$, and $z_3$ become arbitrarily small while $z_1$ becomes positive definite, so there will exist a range of positive $z$ for which $g_{t t}$ is positive.   Likewise, if $k$ becomes large, $Q_0 \sim 1$ and $z_0$ becomes small while $z_1 \sim a^4 g^4$ and hence for sufficiently small $z$ there will be some region where $g_{t t}$ becomes positive.  Examining the cubic roots of the numerator of $g_{t t}$ for the case $e_0 = g/{\sqrt{3}}$ one finds there are no ergospheres if and only if $p = 1$ and $k = 1$ or $p = 1$ and $k = 2$.   In these cases two roots of the cubic are complex and the other root is negative.  Plotting the other roots, noting that due to (\ref{EqDirac3}) since $Q_0 < 1$ that if $p = 1$, $\vert a g \vert > 0.6180$ and if $p \geq 2$, $\vert a g \vert > 0.7807$, one always finds two positive roots for $z$ and hence the double ergosphere structure we have described.  See Figure \ref{EqRoterg} for $g_{t t}$ for the first several cases for values of $p$ and $k$ with $e_0 = g/{\sqrt{3}}$.\footnote{The above simple observations and examples contradict the statement of \cite{RossBubbles} regarding an ergoregion; the analysis there is rather more complicated in that coordinate system and we suspect an algebraic error has occurred.}
	
	\subsection{BPS bound}
	
\begin{figure}[t]

\centering
	\includegraphics[scale= .6]{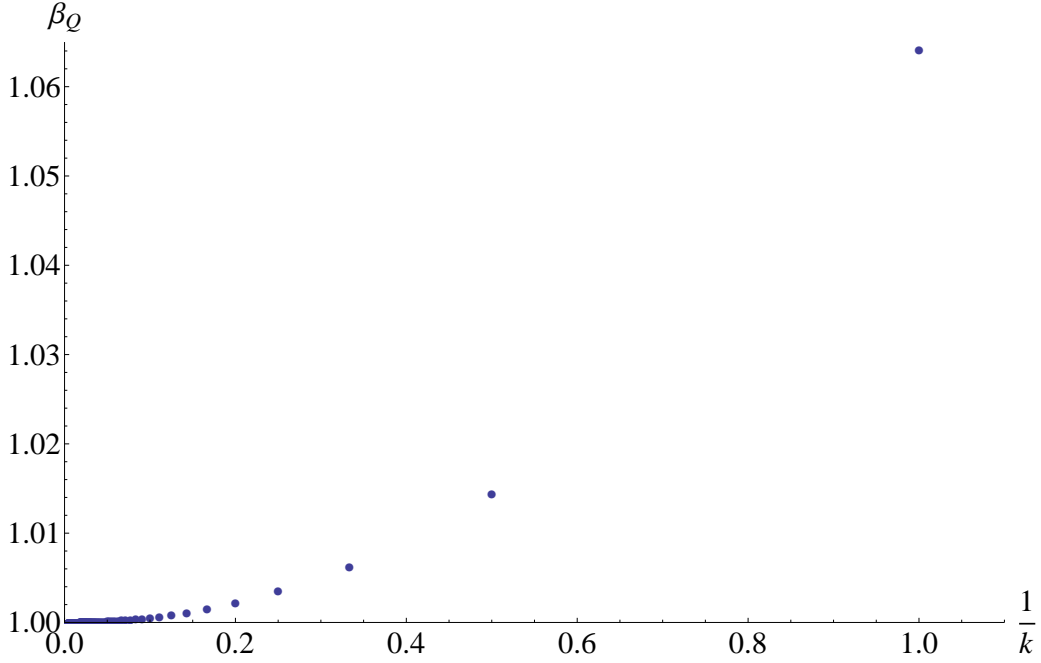}
	\caption{BPS violation for the equally rotating asymptotically flat soliton versus orbifold integer $k$ ($k = 1$: no orbifold, $k \geq 2$: $\mathbb Z_k$ orbifold)}
	\label{CaseIplot2}
	\end{figure}

	Now turning to the BPS bound, for the asymptotically flat equally rotating case
	\be
	\beta_Q = \frac{\vert q \vert}{m} = \frac{2}{1+\sqrt{Q_0}}
	\ee
	which is manifestly larger than one since $0 < Q_0 < 1$.  Specifically for the equally rotating asymptotically flat case one finds
	\be
	\beta_Q = \sec \Bigg[\frac{1}{3} \arccos \Big[1 - \frac{1}{2 k^2}\Big] \Bigg]
	\ee
As expected from the usual observation that deficit angles correspond to positive mass contributions \cite{HawkingHorowitz}, this is maximized at $k = 1$ in which case $\beta_Q \approx 1.064$.  See Figure \ref{CaseIplot2}.

For the asymptotically AdS case,
\be
\beta_Q 
= 1 + \frac{(1- \vert a g \vert ) \Big[ 3 (1 - \sqrt{Q_0})+\vert a g \vert ( 3 + \sqrt{Q_0} + \vert a g \vert + a^2 g^2)\Big]}{(1-a^2 g^2) (3 - a^2 g^2) + \sqrt{Q_0} (3 + a^2 g^2)}
\ee
Note then $\beta_Q$ is strictly greater than one and we again violate the BPS bound.\footnote{This solution was previously discussed in \cite{RossBubbles}, but the notion of the BPS bound applied there is $m  -  \vert q \vert > 0$, which corresponds to the usual notion of the bound for asymptotically flat spaces but not asymptotically AdS ones.}  
	\begin{figure} [t]
\centering
\resizebox{\textwidth}{!}{\mbox{\includegraphics{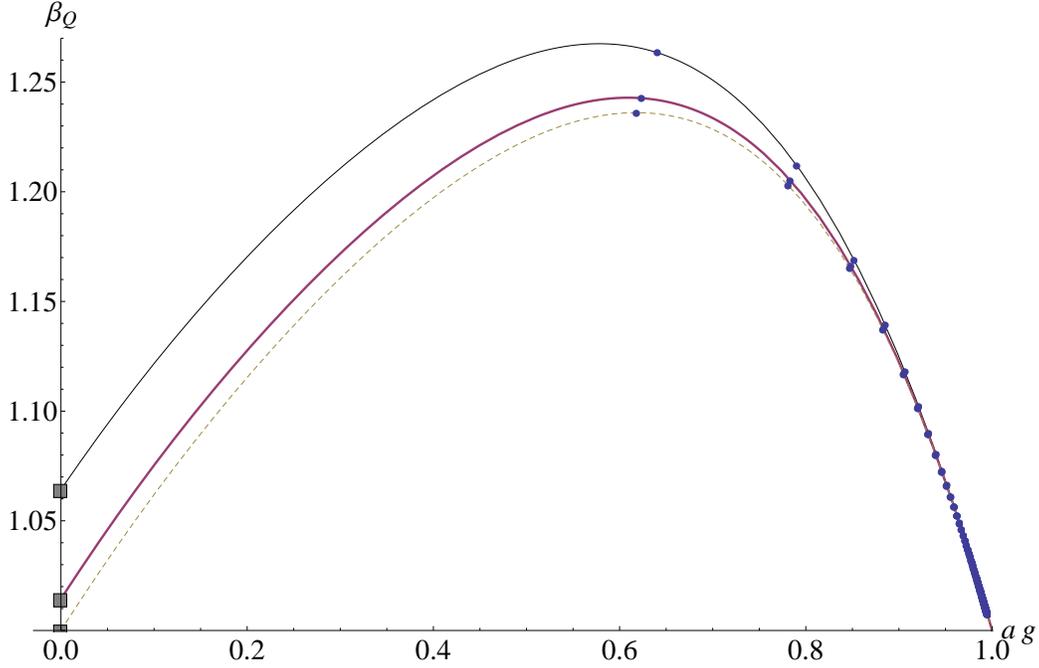}}}
	\caption{BPS violation for the equally rotating asymptotically AdS solutions as a function of the dimensionless rotation $a g < 1 $. The upper curve corresponds to $k = 1$, the second curve to $k = 2$, and the lowest curve to $k \rightarrow \infty$. For any fixed $e_0$, only a discrete family of points along the plotted curves are actually realized; circles mark these points for $p$ between 1 and 100 for $e_0 = g/{\sqrt{3}}$.  The distinct class of asymptotically flat solitons with $g = 0$ are shown with square marks.} 
	\label{ViolBPSEqual}
	\end{figure}
See Table 1 and Figure 4 for some specific examples.  The violation of the bound is persistent, but relatively modest.  Consider in some detail the case $e_0 = g/{\sqrt{3}}$.  Just from the fact that $k \geq 1$ and $p \geq 1$ one can show with a few numerical plots that $\vert a g \vert > 0.618$ and $Q_0 > 0.842$ and then  $\beta_Q < 1.267$.  Hence, the bubble with $p = k = 1$ is very close to the maximum possible value of $\beta_Q$, although verifying the intuition that it will correspond to the maximum possible value of $\beta_Q$ for a smooth solution appears to be technically nontrivial.   The fact that $\beta_Q$ is bounded above is not terribly surprising; if it could be made parametrically large one might expect to obtain a rapidly expanding bubble, not a stationary solution.   It is somewhat surprising that $\beta_Q$ can be larger in the asymptotically AdS case than in the asymptotically flat case, but the bound for AdS includes contributions from the angular momentum absent in the asymptotically flat case.

\begin{table}[t]
\begin{center}
\begin{tabular}[c]{| c | c | | c | c|| c | c|}
\hline
(p,k) & $\beta_Q$ & (p,k) & $\beta_Q$  & (p,k) & $\beta_Q$ \\
\hline
(1,1) & 1.2634 & (1,2) & 1.2427 & (1,10) & 1.2363\\
(2,1) & 1.2121& (2,2) & 1.2053 & (2,10) & 1.2032 \\
(3,1) & 1.1691 & (3,2) & 1.1662 & (3,10) & 1.1653 \\
(4,1) & 1.1392 & (4,2) & 1.1377 & (4,10) & 1.1373 \\
(5,1) & 1.1180 & (5,2) & 1.1171 & (5,10) & 1.1168\\
(10, 1) & 1.0664 & (10,2) & 1.0662 & (10,10) & 1.0662 \\
(25,1) & 1.0286 & (25,2) & 1.0285 & (25,10) & 1.0285\\
(100, 1) & 1.0074 & (100,2) & 1.0074 & (100,10) & 1.0074\\
\hline
\end{tabular}
\end{center}
\caption{BPS violation for equally rotating AdS solitons with $e_0 = g/{\sqrt{3}}$}
	\label{EqrotAdSBPS}
	\end{table}

\setcounter{equation}{0}
\section{Asymptotically flat unequal angular momenta}

\subsection{Smoothness}

In this case we will solve for Q in terms of $\gamma_0$ (\ref{SC1}) and use the conical singularity condition to solve for $b/a$.  As in the previous asymptotically flat case, the Dirac quantization condition will merely quantize the overall dimensionful scale (say $a$).
If one defines
\be
x_0 = \sqrt{(a^2-b^2)^2 + 4 a^2 b^2 Q}
\ee
and
\be
b_0 = \frac{b}{a}
\ee
then if $b^2 \neq a^2$ (\ref{SC1}) can be solved only if
\be \label{IIa}
b_0 \gamma_0 > 0
\ee
and further
\be \label{IIb}
b_0 \gamma_0 \neq 1
\ee

Imposing these restrictions one finds from (\ref{SC1})
\be
x_0 =  \frac{ a^2 (b_0^2-1) (b_0 \gamma_0 + 1)}{1 - b_0 \gamma_0}
\ee
and further
\be \label{IIQ}
Q = \frac{ (b_0^2-1)^2}{(1 - b_0 \gamma_0)^2} \frac{\gamma_0}{b_0}
\ee
Note we have two non-trivial constraints we must impose for the above to make sense, namely $x_0 > 0$ and $ 0 < Q < 1$.  If $\gamma_0^2 = 1$ one quickly finds that $Q > 1$, so we must forbid that case.  The necessary and sufficient condition for all of the above requirements to be true is that if $\vert \gamma_0 \vert  < 1$
\be \label{IIc}
1 < \vert b_0 \vert  < \vert \gamma_0\vert ^{-1/3}
\ee
and if $\vert \gamma_0 \vert > 1$ then
\be \label{IId}
 \vert \gamma_0\vert ^{-1/3}< \vert b_0 \vert < 1
\ee
provided one chooses the sign of $b_0$ such that $b_0 \gamma_0 > 0$.  

The absence of a conical singularity (\ref{SC2}) in this case becomes
\be \label{IIe}
\frac{(1 - b_0^3 \gamma_0) (1 - \gamma_0^2)^2}{(1 - b_0 \gamma_0)^3} = \frac{1}{m_0^2 k^2}
\ee
and since this is a cubic in $b_0$ we may solve it explicitly.  In all cases, it turns out there is only one relevant root.

For the special case where  $m_1 = \pm (m_0 + 1)$ and $k = 1$ the single root such that $ b_0 \gamma_0 > 0$ is
\be \label{IIf}
b_0 = \pm \frac{3 m_0 + 1}{3 m_0 + 2}
\ee
and likewise in the case $m_1 = \pm (m_0 -1)$ (and taking $m_0 \geq 2$ since we forbid $\gamma = \alpha = 0$) and $k = 1$ the single root such that $b_0 \gamma_0 > 0$ is
\be \label{IIg}
b_0 = \pm \frac{3 m_0 - 1}{3 m_0 - 2}
\ee
One can show that (\ref{IIf}) and (\ref{IIg}) satisfy (\ref{IId}) and (\ref{IIc}), respectively, so we have met our smoothness conditions.

Now turning to the more generic case, let us define
\be
C = \frac{m_0^2 k^2 \gamma_0^2 (\gamma_0^2-1)^3 [ m_0^4 k^4 (\gamma_0^2-1)^3 + m_0^2 k^2 (1 + \gamma_0^2 - 2 \gamma_0^4) + \gamma_0^2]}{2 \Big[ m_0^2 k^2 (\gamma_0^2-1)^2-\gamma_0^2 \Big]^3}
\ee
and
\be
D = \frac{m_0^4 k^4 \gamma_0^4 (\gamma_0^2 - 1)^6 [ m_0^4 k^4 (\gamma_0^2-1)^2- 2 m_0^2 k^2(\gamma_0^2+1) +1]}{4 \Big[m_0^2 k^2 (\gamma_0^2-1)^2-\gamma_0^2 \Big]^4}
\ee
It is straightforward to check that the denominators of $C$ and $D$ are never vanishing (recalling that $\gamma_0 = m_1/m_0$ and $m_1$ and $m_0$ are both integers), so the above expressions are sensible.  Except for the special cases  (\ref{IIf}) and (\ref{IIg}), one can show $D > 0$ and furthermore $C - \sqrt{D} > 0$.  Then the single real root of the cubic is
\be \label{CaseIIroot}
b_0 \gamma_0 = - \frac{\gamma_0^2}{m_0^2 k^2  (\gamma_0^2-1)^2-\gamma_0^2} + \Big(C + \sqrt{D}\Big)^{1/3}+ \Big(C - \sqrt{D}\Big)^{1/3}
\ee
There is no obvious analytic method to determine whether (\ref{CaseIIroot}) falls in the relevant ranges for smoothness (\ref{IIc}, \ref{IId}), but where numerics are reliable they consistently show the roots respect these bounds.  The caution in the prior statement is due to the fact that one loses numerical control (at least in the simple approach we have used) as $\vert \gamma_0 \vert \rightarrow \infty$ or $m_0^2 k^2 \rightarrow \infty$.  The appropriate asymptotic series, however, show no signs of any difficulties so we believe (\ref{CaseIIroot}) always obeys the appropriate smoothness bounds.

\subsection{BPS bound}

In the asymptotically flat case the ratio of charge to mass is
\be
\beta_q = \frac{\vert q \vert}{m} = \frac{2 (1 - b_0 \gamma_0)}{\vert b_0 \vert - \vert \gamma_0\vert }
\ee
once we impose the value of $Q$ (\ref{IIQ}) and take care to choose signs according to $b_0$ in the appropriate ranges (\ref{IIc}, \ref{IId}).   First let us consider whether there are any solitons which saturate the BPS bound.  $\beta_q$ will equal 1 if (and only if)
\be \label{BPSIIa}
\vert b_0 \vert = \frac{2 + \vert \gamma_0 \vert }{1 + 2 \vert \gamma_0 \vert}
\ee
and further it is straightforward to check (\ref{IIc}) and (\ref{IId}) are obeyed for the appropriate ranges of $\gamma_0$.   The absence of a conical singularity (\ref{IIe}) becomes
\be
(1 - \vert \gamma_0 \vert )^2 = \frac{1}{m_0^2 k^2}
\ee
or equivalently
\be \label{BPSIIb}
(m_0 - \vert m_1\vert)^2 = \frac{1}{k^2}
\ee
(\ref{BPSIIb}) will then be satisfied if and only if $k = 1$ and $\vert m_1 \vert = m_0 \pm 1$.   These are precisely the special cases we described in the section above (\ref{IIf}, \ref{IIg}).  Under normal circumstances the saturation of the BPS bound would automatically imply these solutions are supersymmetric.  However, since we know the bound can be violated this usual conclusion is subject to question; it would be interesting to explicitly check for the existence of Killing spinors.  To the best of our knowledge, this particular class of solutions has not been noticed before.

For the remaining cases we must require that $b_0$ is given by (\ref{CaseIIroot}) to avoid a conical singularity at the bubble.  While it is difficult to make any analytic statements about $\beta_q$ once this is imposed, we find numerically the BPS bound is always respected in this case.  As before, this conclusion is subject to a caveat that one loses numerical control as $m_0 k \rightarrow \infty$ and $\gamma_0 \rightarrow \infty$, although again expansions around those points give results respecting the BPS bound.  Intuitively, the above results reflect the fact that one expects the most severe violations of the bound to be in the most symmetric situation ($\gamma_0 = 1$), as usually such configurations minimize the mass.  Taking the values of $\gamma_0$ as close to one as possible results in BPS saturating solitons and any other choice gives solitons satisfying the bound.  Alternatively, one can understand the increase in mass, and decrease in $\beta_Q$, as the result of the increased orbifolding one must perform as the angular momenta become more unequal.

\subsection{Ergoregion}

In this case the norm of $\partial/{\partial t}$ is given by
\be
g_{t t} = -\frac{\rho^4 - 2 m \rho^2 + q^2}{\rho^4}
\ee
and so there are two surfaces were $g_{t t}$ vanishes
\be
\rho_{\pm}^2 = m \pm \sqrt{m^2 - q^2}
\ee
Since $\rho^2 = r^2 + a^2 \cos^2 \theta + b^2 \sin^2 \theta$, the maximum and minimum value of $\rho_{\pm}^2$ occur at $\theta = 0$ and $\theta = \pi/2$, although which is a maxima and which a minima depends on the relative magnitude of $a$ and $b$.  Recalling that $m^2 - q^2$ is non-negative if and only if the BPS bound is satisfied (\ref{flatBPS1}), since all the solutions in the present class respect this bound we will have ergosurfaces provided $\rho_{\pm}^2$ is large enough to be outside the bubble.  Further, if $\rho_{-}^2$ is outside the bubble we will have inner and outer ergosurfaces.   In the special case of BPS saturating solitons (\ref{IIf}, \ref{IIg}), the inner and outer ergosurfaces are at the same radius and $g_{t t}$ has a second order zero:
\be
g_{t t} = -\Big(1 - \frac{m}{\rho^2}\Big)^2
\ee
It will be handy to define the radial location of the ergosurfaces as $R_{\pm}(\theta)$ 
\be
R_{\pm}(\theta)= \rho_{\pm}^2 - a^2 \cos^2 \theta - b^2 \sin^2 \theta =  m \pm \sqrt{m^2 - q^2}- a^2 \cos^2 \theta - b^2 \sin^2 \theta 
\ee

\begin{figure}[t]
	\begin{picture} (0,0)
    	\put(18,7){$z$}
         \put(386, -216){$\theta$}
    \end{picture}

\centering
	\includegraphics[scale= 1.05]{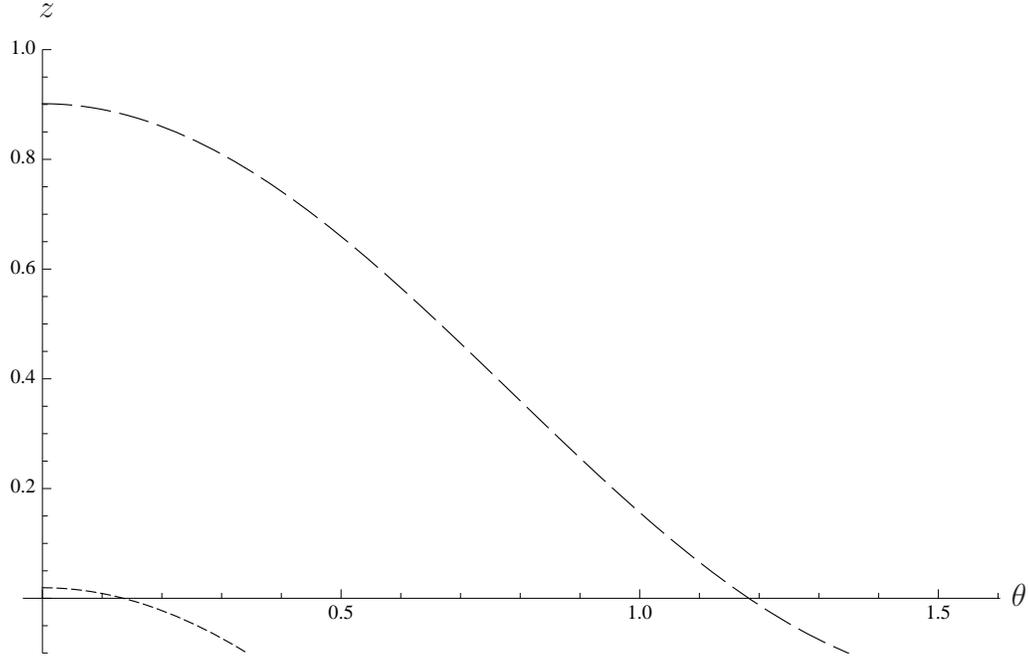}
	\caption{Inner [short dashes] and outer [long dashes] ergosurfaces for $\gamma_0 = 1/3$, $k = 1$, where $z$ parametrizes the radial distance from the bubble.} 
		\label{ergpict}
	\end{figure}

Via a few graphical plots one can show that if $b^2 < a^2$ ($\gamma_0^2 > 1$), including the BPS saturating soliton with $k= 1$ and $\vert m_1 \vert = m_0 + 1$,
\be \label{AFeqn1}
R_{\pm}(0) < R_0
\ee
and
\be \label{AFeqn2}
R_{\pm} (\pi/2) > R_0
\ee
or in other words, in traveling along the axis $\theta = 0$ one encounters no ergosurface before encountering the bubble but along the axis $\theta = \pi/2$ once goes through two ergosurfaces before reaching the bubble. Likewise, if $b^2 > a^2$ ($\gamma_0^2 < 1$), including the BPS saturating soliton with $k = 1$ and $\vert m_1 \vert = m_0 - 1$,
\be
R_{\pm}(0) > R_0
\ee
and
\be
R_{\pm} (\pi/2) < R_0
\ee
and the situation is reversed from that above, as one would have predicted via symmetry.  Defining a radial distance $z$ via $R_{\pm} = R_0 +  a^2\, z$, we have plotted an example of these surfaces in Figure \ref{ergpict}.  Note that each of these ergosurfaces runs into the bubble at a finite value of $\theta$ between 0 and $\pi/2$. However, since the bubble is a place where a cycle smoothly degenerates these ergosurfaces are, in fact, manifolds without boundaries (radially incoming geodesics are reflected back outwards at a shifted value of $\psi$) .

In fact these ergosurfaces, which we will refer to as capping spheres, surround a pole of the minimal $S_2$ (i.e. the bubble) and are topologically spheres.   To show this, let us first note these surfaces are simply connected.  One may describe a round $S_3$
\be
ds^2 = d \theta^2 + \sin^2 \theta d \phi_1^2 + \cos^2 \theta d \phi_2^2
\ee
by a sequence of tori at fixed values of $\theta$ ranging between $0$ and $\pi/2$.  The two directions of the torus at any fixed $\theta$ are parametrized by $\phi_1$ and $\phi_2$ and the tori degenerate into circles at $\theta = 0$ and $\theta = \pi/2$.   See Figure {\ref{Tor1}. 

\begin{figure}[h]
	\begin{picture} (0,0)
    	\put(-8,-70){$\theta = 0$}
	\put(130,-70){$\theta = \pi/4$}
         \put(335, -70){$\theta = \pi/2$}
    \end{picture}

\centering
	\includegraphics[scale= .75]{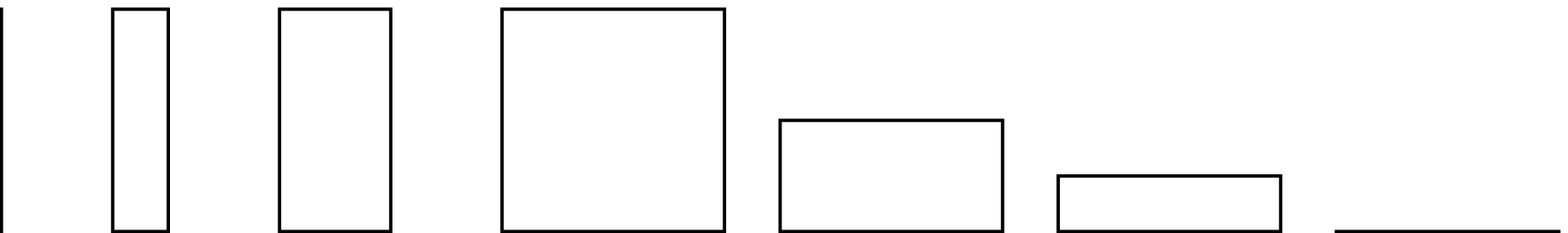}
	\bigskip
	\caption{$S_3$ as a sequence of tori} 
		\label{Tor1}
	\end{figure}

For these capping sphere ergosurfaces we do not obtain a complete series of tori but rather ones only up to some particular value of $\theta$, say $\theta_0$, where we run into the bubble.  For the sake of definiteness consider ergosurfaces of the type (\ref{AFeqn2}) which surround the pole at $\theta = 0$ and run into the bubble at $\theta_0 < \pi/2$.   If we simply cutoff the series of tori at $\theta_0$ we would be left with an apparently non-contractible cycle  along the $\phi_2$ direction.  However, as we approach $\theta_0$ the two directions $\phi_1$ and $\phi_2$ degenerate and point in the same  direction (more precisely, each of $\phi_1$ and $\phi_2$ can be seen as a combination of the angles $\psi$ and $\phi$ and as $\psi$ pinches off the surviving portions of $\phi_1$ and $\phi_2$ both point in the $\phi$ direction).  In terms of the sequence of tori this means the angle between the $\phi_1$ and $\phi_2$ sides is going to zero (one has an increasingly narrow parallelogram) and at $\theta_0$ the two sides touch, as shown in Figure \ref{Tor2}.

	\begin{figure}[b]
	\begin{picture} (0,0)
    	\put(-10,-70){$\theta = 0$}
         \put(347, -70){$\theta = \theta_0$}
    \end{picture}

\centering
	\includegraphics[scale= .7]{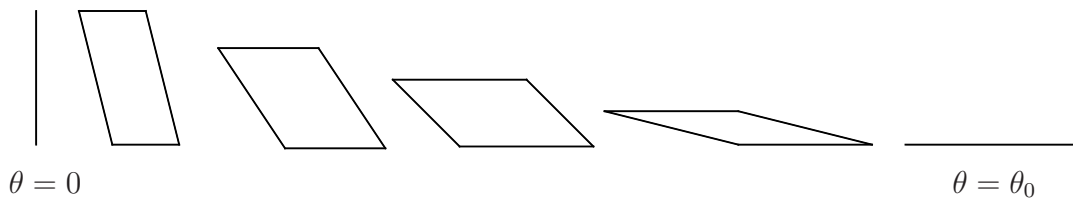}
	\bigskip
	\caption{Sequence of squashed tori in the capping ergosphere} 
		\label{Tor2}
	\end{figure}

Now let us consider whether there are any non-contractible cycles in this ergosurface.  Cycles along the $\phi_1$ direction may be contracted to a point by dragging the curve to $\theta = 0$ where $\partial/{\partial \phi_1}$ degenerates.   Cycles along the $\phi_2$ direction may be dragged to $\theta = \theta_0$, at which point the portion of the cycle which does not degenerate is also lying along the $\phi_1$ direction.   This remaining cycle may then be dragged to $\theta = 0$ and shrunk to a point.   A generic curve, viewed as a combination of cycles in the $\phi_1$ and $\phi_2$ direction, may be contracted by first removing the $\phi_1$ portion by going to $\theta = 0$ and then going to $\theta = \theta_0$ and repeating the above steps to remove the $\phi_2$ portion.   Hence all cycles in the ergosurface are contractible.   Then this ergosurface is a compact simply connected manifold without boundary; via the Poincar\'e conjecture such a manifold is topologically equivalent to a sphere. 

\setcounter{equation}{0}
\section{Unequal angular momenta asymptotically AdS}

\subsection{Smoothness}

In this case we will solve again solve for $Q$ in terms of $\gamma_0$ (\ref{SC1}).  We will be able to do so only if
\be \label{IIIa}
b_0 \gamma_0 > 0
\ee
and further
\be \label{IIIb}
b_0 \gamma_0 \neq \frac{1 - b^2 g^2}{1-a^2 g^2}
\ee
Given these restrictions, from (\ref{SC1}) 
\be \label{IIIx}
x_0 =  \sqrt{(a^2-b^2)^2 + 4 a^2 b^2 Q} = \frac{a^2 (b_0^2 - 1) [ 1 - b_0^2 a^2 g^2 + b_0 \gamma_0 (1 - a^2 g^2)]}{1 - b_0^2 a^2 g^2 - b_0 \gamma_0 (1 - a^2 g^2)}
\ee
Note the term in brackets in the numerator of (\ref{IIIx}) is positive definite given (\ref{IIIa}) and the conditions that $a^2 g^2 < 1$ and $b_0^2 a^2 g^2 < 1$ ($\Xi_a > 0$ and $\Xi_b > 0$).   Then the  condition that $x_0 > 0$ is
\be \label{xconst1}
1 - b_0^2 a^2 g^2 - b_0 \gamma_0 (1 - a^2 g^2) < 0
\ee
if $b_0^2 < 1$ and
\be \label{xconst2}
1 - b_0^2 a^2 g^2 - b_0 \gamma_0 (1 - a^2 g^2) > 0
\ee
if $b_0^2 > 1$.   Then $Q$ becomes
\be
Q = \frac{(b_0^2-1)^2 (1 - a^2 g^2) (1 - b_0^2 a^2 g^2)}{[1-b_0^2 a^2 g^2 - b_0 \gamma_0 (1 - a^2 g^2)]^2} \frac{\gamma_0}{b_0}
\ee
The statements that $0 < Q < 1$, $\Xi_a > 0$, $\Xi_b > 0$, as well as (\ref{IIIb}), (\ref{xconst1}) and  (\ref{xconst2}), are equivalent to the requirement
\be \label{AdSsmooth1}
0 < a^2 g^2 < \frac{ b_0^3 \gamma_0 -1}{b_0^2 (b_0 \gamma_0 - 1)}
\ee
and
\be \label{AdSsmooth2}
b_0 \gamma_0 > 0
\ee
and if $\vert \gamma_0 \vert < 1 $ then
\be \label{AdSsmooth3}
1 < \vert b_0 \vert <\vert \gamma_0 \vert^{-1/3}
\ee
and if $\vert \gamma_0 \vert > 1 $
\be \label{AdSsmooth4}
\vert \gamma_0 \vert^{-1/3} < \vert b_0 \vert < 1
\ee
We note (\ref{AdSsmooth2}-\ref{AdSsmooth4}) are the same requirements we had in the asymptotically flat case and (\ref{AdSsmooth1}) becomes trivial as $g \rightarrow 0$.

While the absence of a conical singularity (\ref{SC2}) may be regarded as as cubic equation for $a^2 g^2$, there does not seem to be any simple description of the subsequent roots.  Hence, we will at this point impose the Dirac quantization condition with charge $e_0 = g/\sqrt{3}$, suitable for embedding in ten dimensional supergravity as discussed above, which becomes the statement that
\be \label{Dirac3}
\frac{a g (b_0^2 - 1)}{1-b_0 \gamma_0 + b_0 (\gamma_0 - b_0) a^2 g^2} = p \, m_0
\ee
for some integer $p$.  Note from the form of (\ref{Dirac3}), together with the above requirements (\ref{xconst1}) and (\ref{xconst2}), implies we must choose signs such that $p a g > 0$.   This then implies that if $\vert \gamma_0 \vert < 1$ the single suitable root of $a g$ from (\ref{Dirac3}) is
\be \label{Dirac4}
a g = \frac{1-b_0^2 + \sqrt{(1-b_0^2)^2 + 4 p^2 m_0^2 b_0 (1 - b_0 \gamma_0 ) (b_0-\gamma_0)}}{2 p m_0 \, b_0 (b_0 - \gamma_0)}
\ee
and if $\vert \gamma_0 \vert > 1$ it must be true that
\be \label{Dirac5}
a g = \frac{b_0^2-1  + \sqrt{(1-b_0^2)^2 + 4 p^2 m_0^2 b_0 (1 - b_0 \gamma_0 ) (b_0-\gamma_0)}}{2 p m_0 \, b_0 (\gamma_0 - b_0)}
\ee
The absence of a conical singularity at the bubble (\ref{SC2}) becomes
\be \label{SC4}
\frac{[ 1 - b_0^2 a^2 g^2 - b_0^3 \gamma_0 (1 - a^2 g^2) ] [1-b_0^2 a^2 g^2 - \gamma_0^2  (1 - a^2 g^2)]^2}{[1-b_0^2 a^2 g^2 - b_0 \gamma_0 (1 - a^2 g^2)]^3} = \frac{1}{m_0^2 k ^2}
\ee
and may be regarded as determining $b_0$ in terms of the integers $p$, $m_1$, $m_0$, and $k$ (given (\ref{Dirac4}) or (\ref{Dirac5}) as appropriate).  There does not appear to be any obvious way to make any analytic statements for this case.  Even numerically thoroughly exploring this four parameter space would be a nontrivial exercise, so for the present we have contented ourselves with examining a variety of numerical examples which we hope are not atypical.   In each of these examples, numbering more than a hundred, (\ref{SC4}) has a single root for $\vert b_0 \vert$ in the bounds (\ref{AdSsmooth3}) and (\ref{AdSsmooth4}) and (\ref{AdSsmooth1}) is always satisfied.  In particular the examples we later list when examining ergoregions and the BPS bound satisfy these limits.

\subsection{BPS bound}

The complicated smoothness conditions above have prevented us from making any analytic analysis regarding the BPS bound in the unequally rotating AdS case, but remarkably enough in the many examples we have examined the BPS bound is always violated. 
\begin{table}[t]
\begin{center}
\begin{tabular}[c]{|c | c | c | |c | c | c ||c | c | c | }
\hline
$m_1$ & $m_0$ & $\beta_Q$ &$m_1$ & $m_0$ & $\beta_Q$ & $m_1$ & $m_0$ & $\beta_Q$\\
\hline
1& 100 & 1.0114 & 5 & 9 & 1.0890 &2 & 1& 1.2210 \\
2& 51 & 1.0241 & 3& 5 & 1.1361 & 11  & 4 & 1.0830 \\
1& 4 & 1.1629 & 87 & 121 & 1.0071 &9 & 2 & 1.1017 \\
3& 10 & 1.0921 & 3 & 4 & 1.1498 &23 & 3 & 1.0495 \\
1& 3 & 1.1880 & 4 & 5 & 1.1261 &27 & 1 & 1.0390 \\
2& 5 & 1.1452 & 9& 10 & 1.0692 &1354& 19 & 1.0011 \\
3& 7 & 1.1145  & 93 & 101 & 1.0076 &100& 1 & 1.0114 \\
1& 2 & 1.2210 & 113 & 117 & 1.0065 &538 & 1 & 1.0022 \\
\hline
\end{tabular}
\end{center}
\caption{Examples of BPS violation for $g \neq 0$  solitons  with $p = k = 1$}
	\label{AdSBPSI}
	\end{table}
	\begin{table}
\begin{center}
\begin{tabular}[c]{| c | c | | c | c|| c | c|}
\hline
(p,k) & $\beta_Q$ & (p,k) & $\beta_Q$  & (p,k) & $\beta_Q$ \\
\hline
(1,1) & 1.2210 & (1,5) & 1.2139 & (1,100) & 1.2136\\
(2,1) & 1.1634 & (2,5) & 1.1615 & (2,100) & 1.1614 \\
(3,1) & 1.1251 & (3,5) & 1.1244 & (3,100) & 1.1244 \\
(4,1) & 1.1007 & (4,5) & 1.1003 & (4,100) & 1.1003 \\
(5,1) & 1.0841& (5,5) & 1.0839 & (5,100) & 1.0839\\
(10, 1) & 1.0458 & (10,5) & 1.0458 & (10,100) & 1.0458 \\
(25,1) & 1.0193 & (25,5) & 1.0193 & (25,100) & 1.0193\\
(100, 1) & 1.0050 & (100,5) & 1.0050 & (100,100) & 1.0050\\
\hline
\end{tabular}
\end{center}
\caption{Examples of BPS violation for $g \neq 0$  solitons  with $\gamma_0 = 1/2$}
	\label{AdSBPSII}
\end{table}
See Table 2 for $p = k = 1$ with various values of $m_1$ and $m_0$ and Table 3 for $m_0 = 1$, $m_1 = 2$ and various values of $(p, k)$.\footnote{Since the violation of the bound is independent of the relative signs of the angular momenta, we just list positive values for $\gamma_0$, but one obtains identical values of $\beta_Q$ for counter-rotating solutions.}   The values listed here do not differ substantially from other cases we have examined and appear to be ``typical'', insofar as we can tell.  As a phenomenological observation, $\beta_Q$ appears to be maximized when all values of the integers $(m_0, m_1, p, k)$ are as small as possible. It becomes  close to (but still greater than) one when any one of integers ($m_0, m_1, p$) becomes large.  $\beta_Q$ appears to be remarkably independent of the orbifolding integer $k$.  For all cases we have examined, $\beta_Q$ has always been less than that of the $\vert p \vert  = k =1$ equally rotating AdS solution.

\subsection{Ergoregion}

If we define a distance away from the bubble $z$ as
\be
z = R - R_0
\ee
then
\be \label{AdSgtt}
g_{t t} = \frac{\Delta_\theta}{\Xi_a^2 \Xi_b^2 \rho^4} (c_3 z^3 + c_2 z^2 + c_1 z + c_0)
\ee
where
\be
c_3 = -g^2 \Xi_a \Xi_b
\ee
\be
c_2 = - \Xi_a \Xi_b \Bigg[ \Delta_\theta + \frac{3 g^2}{2} \Big( (a^2-b^2) \cos 2\theta + \sqrt{(a^2-b^2)^2+ 4 a^2 b ^2 Q} \Big) \Bigg]
\ee
and $c_1$ and $c_0$ are somewhat complicated functions of $\theta$, although easily found given the metric above.   Note $c_3$ and $c_2$ are negative definite.  The signs of $c_1$ and $c_0$ depend on values of the parameters, as well as $\theta$, but note that
$$
c_0 (\theta = 0) = - \frac{\Xi_a \Xi_b^2}{4} \Big(a^2 - b^2 +\sqrt{(a^2-b^2)^2 + 4 a^2 b^2 Q} \Big) \Big [ a^2 - b^2 -2 a^2 Q 
$$
\be
+ \sqrt{(a^2 - b^2 - 2 a^2 Q)^2 + 4 a^4 Q (1 - Q)} \Big]
\ee
and likewise
$$
c_0 (\theta = \pi/2) = -\frac{\Xi_a^2  \Xi_b}{4} \Big(b^2 - a^2 +\sqrt{(b^2-a^2)^2 + 4 a^2 b^2 Q} \Big) \Big [ b^2 - a^2 -2 b^2 Q 
$$
\be
+ \sqrt{(b^2 - a^2 - 2 b^2 Q)^2 + 4 b^4 Q (1 - Q)} \Big]
\ee
Note then that $c_0$ is negative definite at the poles of the bubble ($\theta = 0$, $\theta = \pi/2$).  Thus at the bubble at $\theta = 0$ and $\theta = \pi/2$, $g_{t t}$ is negative definite and we are not inside an ergoregion.  Traveling along the axis $\theta = 0$ and $\theta = \pi/2$ one can encounter either no ergosurface or two ergosurfaces -- there are no solitons with a single ergosurface 
surrounding the entire bubble.

	\begin{figure} [t]
	\begin{picture} (0,0)
    	\put(29,7){$z/l$}
         \put(376, -191){$\theta$}
    \end{picture}

\centering
	\includegraphics[scale=.9]{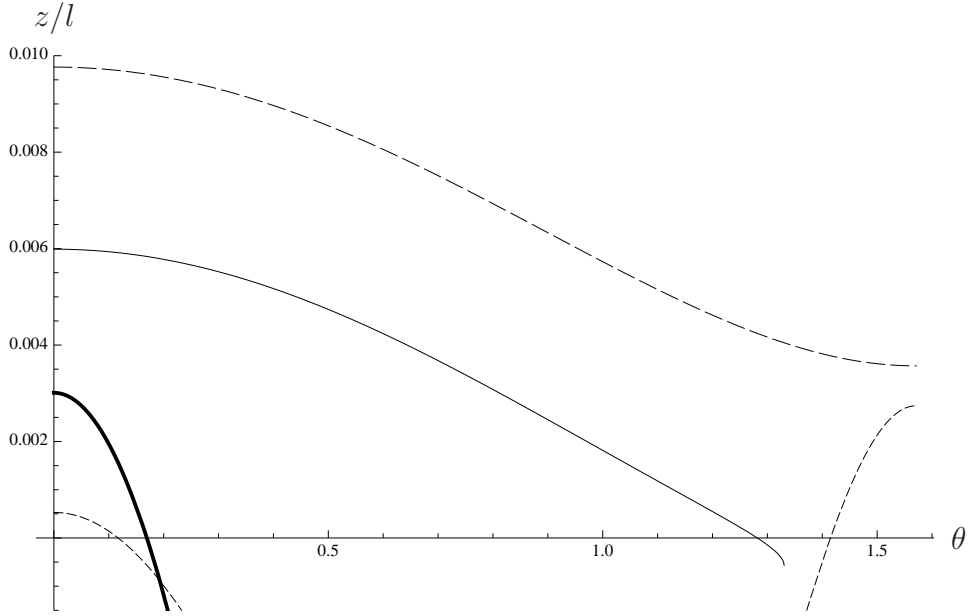}
	\caption{Inner [short dashes] and outer [long dashes] ergosurfaces for $\gamma_0 = 2/3$, $k = p = 1$ and inner [thick line] and outer [thin line] ergosurfaces for $\gamma_0 = 1/2$, $k = p = 1$, where the vertical axis parametrizes the distance from the bubble.}
	\label{AdSergpict}
	\end{figure}

Without imposing the absence of conical singularities and the Dirac quantization condition there seems to be very little one can say beyond the above--plotting the roots to the cubic in (\ref{AdSgtt}) without these extra conditions gives results which are highly parameter dependent and seem to cover all possible ergosurface configurations.  While, as noted above, we do not know how to make any statement for arbitrary values of $m_0$, $m_1$, $p$ and $k$, all of the substantial number of examples we have examined follow a rather simple pattern.  In the case where $m_0 = \vert p \vert = 1$,  we have an ergosurface structure like the asymptotically flat unequally rotating bubbles with double capping spheres  that surround the pole $\theta = 0$ and run into the bubble at values of $\theta$ smaller than $\pi/2$.   Likewise, if $\vert m_1\vert = \vert p \vert = 1$ we find double capping spheres surrounding the $\theta = \pi/2$ axis but no ergoregion around the $\theta = 0$ axis.  For all remaining cases we find a single outer ergosphere surrounds the entire bubble and inner ergosurfaces in the form of capping spheres which surround each pole and do not intersect each other.    Let $z_1$ be the largest (real) root of the cubic (\ref{AdSgtt}) and $z_2$ be the next largest.  We have plotted these $z_i$ (i.e. the location of ergosurfaces) in Figure \ref{AdSergpict} for some particular solitons; note for the sake of visualization we have actually plotted $z_1/{50 l} $ and $z_2/l$.

 \section{Spin structure}
 
If we want to consider fundamental fermions on a background of the type we have described the spin structure must have antiperiodic fermions  around any simple contractible cycle.  In the case of the solitons we have described this then implies fermions are antiperiodic around the $\psi$ direction, since the $\psi$ direction pinches off in the interior of the spacetime.  If $\psi$ was asymptotically a  Kaluza-Klein direction this would then be incompatible with supersymmetric boundary conditions. However, in this case $\psi$ is asymptotically simply part of an $S_3$.  In particular away from the bubble $\psi$ parametrizes a simple closed curve on $S_3$, since any self-intersection would occur only if our map $(\phi_1,  \phi_2) \rightarrow (\psi, \phi)$ were not one-to-one.  Since $S_3$ is simply connected, then by definition this curve is smoothly contractible to an arbitrarily small simple closed curve.  Hence the spin structure will have antiperiodic fermions around this cycle, as well as any other simple closed curve in the $S_3$, and the spin structure may be defined consistently.  Topologically, the modifications we have made to $\psi$ to produce these solitons is the only difference between these solutions and flat space, so the above should be not just necessary but a sufficient check of possible obstructions.\footnote{Unfortunately, the more formal ways one has of verifying that this is a spin manifold do not seem to be practical.  Computing the second Stiefel-Whitney class of a given four-dimensional Riemannian manifold is, to the best of our knowledge, a mathematically nontrivial question and likewise it seems unlikely that one could explicitly solve the Dirac equation on these manifolds.}  

It has been previously asserted \cite{RossBubbles}  that such manifolds are spin only if $m_0 + m_1$ is an odd integer.  Provided one takes care, as we have, to choose cycles so we obtain spacetimes which are globally asymptotically flat or globally AdS (i.e. avoid quotients) all these spacetimes are topologically equivalent, so on general grounds the claim would seem to fail.   Let us now explain in detail why in fact there is not an apparent obstruction to defining a spin manifold.   While the curve we have considered above may go around both the poles $\theta=0$ and $\theta= \pi/2$  multiple times (according to the values of $m_0$ and $m_1$) it is \emph{not} equivalent to a union of simple curves going around  $\theta=0$ and $\theta = \pi/2$ separately.  The latter is topologically inequivalent to the curve we started with (it is self-intersecting and/or disconnected) and the assignment of fermion signs is not continuous under this change of topology, even if one takes care to preserve the orientation of the curve.   

\begin{figure}[t]
    \begin{minipage}[t] {.5\linewidth}
\centering
	\includegraphics[scale= .5]{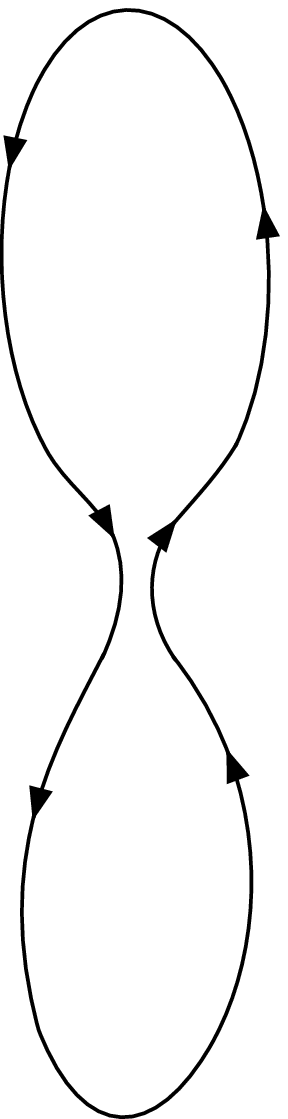}
	\caption {Antiperiodic fermions} 

    \label{figure1}

\end{minipage} %
    \begin{minipage}[t] {.5\linewidth}
    \centering
	\includegraphics[scale=.5]{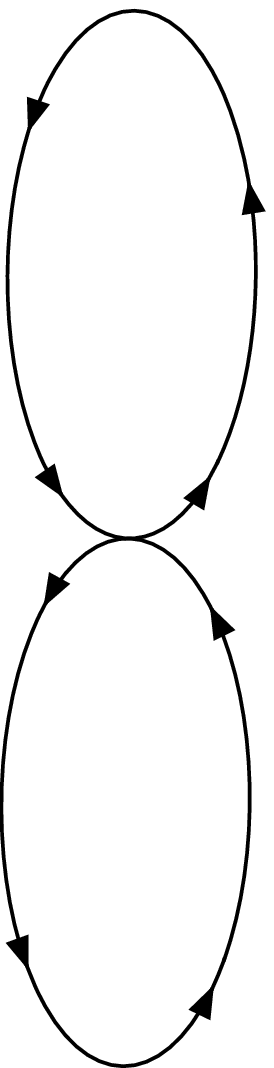}
	\caption{Periodic fermions}

    \end{minipage}
\end{figure}

This latter point is probably most simply illustrated in the plane.  Figure 9 gives a simple contractible curve, with arrows used to indicate an orientation, and upon which the spin structure must have antiperiodic fermions.  Figure 10 illustrates the self-intersecting curve we get if one pinches off the ``neck'' in Figure 9.  Figure 10 is a self-intersecting curve which may be viewed as the union of two simple curves.  Then since the fermions are antiperiodic around each simple curve, they are periodic around the entire self-intersecting curve.  Said another way, propagating a fermion around each simple contractible curve is equivalent to a local Lorentz rotation of $2 \pi$ and propagating around the entire self-intersecting curve equivalent to a local Lorentz rotation of $4 \pi$.   One may consider non-planar curves if one wishes with similar results--given the manifold is simply connected, simple curves in the spin structure must have antiperiodic fermions around them but if one changes the topology of the curve fermions may be periodic or antiperiodic depending on the number of loops (i.e. whether the total curve is equivalent to a local Lorentz rotation that is an integer multiple of $2 \pi$ but not $4 \pi$).

 \section{Summary and Discussion}
 
 To summarize, in the asymptotically flat case, we find equally rotating solitons that violate the BPS bound and are free of ergosurfaces.  There are special classes of asymptotically flat solitons that saturate the BPS bound and have a second order zero in $g_{t t}$ but no finite volume ergoregion.  The remaining asymptotically flat solitons satisfy the bound and  have ergosurfaces we have dubbed capping spheres, which are topologically spheres but run into the bubble.  To our knowledge this kind of ergosurface structure has not been previously observed.
 
 For the asymptotically anti de Sitter solutions, all the smooth solutions we have found violate the BPS bound.  In the equally rotating case one finds an ergoregion disjoint from the bubble surface.   For the unequally rotating Ads solutions, one either has a set of double capping spheres around one pole, or a single outer ergosphere and inner capping spheres around each pole.
 As far as we know, these kinds of structures have never previously been described.
 
 There remains a large class of unanswered questions.  One would like to understand the stability of these solutions.  Due to the absence of horizons, we do not expect any superradiance even in the solitons with ergoregions, but one can argue generically that the presence of any ergoregion signals an instability in the spacetime \cite{ErgInstab}. The issue of stability clearly deserves to be investigated in detail, but note even if some of the solutions are unstable, the violation of the BPS bound (a statement regarding the conserved charges) would be unaffected by such instabilities.  The fact that all the parameters of these solitons end up being quantized is striking.  One would like to know if this is the result of a fundamental limitation or merely due to the family of solutions we have considered, despite its rather universal nature as far as (spherical) black holes is concerned.  On the other hand, it would be interesting to see if one could adapt the uniqueness theorems for spherical black holes for solitons.  For the sake of (relative) simplicity we have focused on five dimensional solutions, but it seems likely that there are analogous solutions in even higher dimensions; in particular one would like to know if there are asymptotically flat solutions of this type in ten or eleven dimensions.

 Perhaps the most striking feature of the solutions we have described above is the violation of the BPS bound.   We should first note that this violation is not necessarily in any sense catastrophic for minimal supergravity.  The solitons all have positive energy and there is no danger they will destabilize the vacuum.  In fact, given that objects which are well described by supergravity tend to have a great deal of entropy (e.g. black holes) it is not clear any objects would decay to these solitons.
 
 On the other hand, the BPS bounds we have mentioned are derived using spinorial proofs following Witten's positive energy theorem and it is important to understand how and why we are violating them.  In the asymptotically AdS case  \cite{AdSBPS} the supercovariant derivative involves a term proportional to the potential and hence it is conceivable that one might produce another term in the theorem proportional to the local magnetic charge, just as one does in a Hamiltonian derivation of the first law of black hole mechanics in the presence of such charges \cite{CopseyHorowitz}.  The asymptotically flat construction, however, appears to be unhampered by the existence of a potential which is globally not defined.  It would seem a $g \rightarrow 0$ limit of even a modified AdS BPS bound should reproduce the asymptotically flat case, as is true for the unmodified versions.   Further, it would be surprising if such a contribution came in with the opposite sign to electric charge to reduce the violation of the bound; usually different charges in a configuration do not act to cancel each other.
 
 There two possible technical obstructions to the implementation of the above theorems.  It is necessary in these arguments to insist that the Dirac equation projected into a spacelike slice vanishes and we are not aware of a demonstration one can always do this in the presence of matter fields and nontrivial topology in more than three spatial dimensions.  Furthermore, one needs to assume the spinors used are asymptotically covariantly constant.  It is conceivable that one can define spinors on this spacetime but they cannot be made regular at the bubble and constant at infinity at the same time.  At the present we have not investigated either one of these possibilities in detail and the apparent contradiction between the solitons and the theorems remains a mystery.
 
With regards to AdS/CFT, this violation of the bound seems particularly surprising.  In the CFT, the bound is simply derivable from the supersymmetry algebra.  We should emphasize the falloff rates of the metric and field are the usual ones (and the same as those for black holes, including the supersymmetric ones, in minimal supergravity) so it seems hopeless to suggest this is some non-normalizable deformation of the usual theory.    It is true, however, the potential is not globally well-defined and presumably the field theory is aware of this.   One might hope that this modifies the supersymmetry algebra, as is known to occur in field theories with topological charges \cite{WittenOlive}.  One might also hope that due to some subtlety, one is allowed to only consider some subset of the eigenvalues of the matrix of $\{Q, Q\}$ and the absolute values normally in the BPS bound are incorrect.  We have the freedom to choose the signs of the electric charge and angular momenta by choosing $a$, $b$, and $s_\epsilon$ appropriately, so any such alternative definition of the bound would somehow have to neutralize this apparent freedom.  We do not know that this is possible.
 
 If none of the above hopes or speculations turns out to be true, one would seem to have a significant mystery for AdS/CFT.   These solitons seem to be perfectly regular and it would seem there is no obstruction, at least in principle, to doing perturbative string theory using them as backgrounds.  One would seem to either have to explain there is some hidden pathology we have missed or explain why it is consistent to ignore such geometries.   Presuming one can overcome the mystery regarding the BPS bound, one would have objects with local charge and unusual ergosurfaces, uncomplicated by the presence of horizons, as new backgrounds to explore the correspondence.
 
\vskip 5cm
\centerline{\bf Acknowledgments}
\vskip .5cm

It is a pleasure to thank  S.Ross, D. Morrison, M.Cvetic,S. Detournay,  D. Marolf, G. T. Horowitz, and H.S. Reall  for useful discussions and correspondence.  We would further like to express our appreciation for hospitality at the Perimeter Institute and the KITP, where a substantial part of this work was completed.  RBM would like to thank the Natural Sciences and Engineering Council of Canada and the Fulbright Foundation for support.  The work of S. de Buyl is funded by the European Commission though the grant PIOF-GA-2008-220338 (Home institution:
Universit\'e Libre de Bruxelles, Service de Physique Th\'eorique et
Math\'ematique, Campus de la Plaine, B-1050 Bruxelles, Belgium).  The work of G.C. is supported in part by the US National Science
Foundation under Grant No.~PHY05-55669 and by funds from the University of
California.

 \appendix
 \setcounter{equation}{0}
 \section{Appendix 1: The Hopf fibration}
 
 Given a three sphere we may parametrize it in the usual way with a polar angle $\theta$ and two periodic directions $\phi_1$ and $\phi_2$
 \be \label{S31}
 ds^2 = d\theta^2 + \sin^2 \theta d \phi_1^2 + \cos^2 \theta d\phi_2^2
 \ee
 where $0 \leq \theta \leq \pi/2$ and $0 \leq \phi_1 < 2 \pi$  and  $0 \leq \phi_2 < 2 \pi$.  One may define combinations of the angles 
 \be
 \psi = \phi_1 + \phi_2
 \ee
 and
 \be
 \phi = \phi_2 - \phi_1
 \ee
 and a rescaled polar angle
 \be
 \bar{\theta} = 2 \theta
 \ee
 so that (\ref{S31}) becomes
 \be
 ds^2 = \frac{1}{4} \Big[ (d\psi + \cos \bar{\theta} \, d \phi)^2 + d\bar{\theta}^2 + \sin^2 \bar{\theta} d \phi^2 \Big]
 \ee
 and we recognize the latter components as the metric on a round $S_2$.   In terms of Cartesian coordinates which cover the sphere we have
 \begin{eqnarray}
x_1 &=& \sin \theta \sin \frac{\psi- \phi}{2} \\
x_2 &= &\sin \theta \cos  \frac{\psi- \phi}{2}\\
x_3 &= &\cos \theta \sin  \frac{\psi+\phi}{2}\\
x_4 &=& \cos \theta \cos  \frac{\psi+ \phi}{2}
\end{eqnarray}
Note that if one insists that one has an entire $S_3$ and $\psi$ and $\phi$ are periodic directions, the periods of $\psi$ and $\phi$ would have to be integer multiples of $4 \pi$.  In particular, under the replacement $\phi \rightarrow \phi + 2 \pi$, $x_i \rightarrow - x_i$.   

On the other hand, if we are to have a diffeomorphism, then the proper area of the sphere in these new coordinates must necessarily be $2 \pi^2$, which implies, after a few lines of algebra, that
\be \label{area2}
\Delta \psi \Delta \phi = 8 \pi^2
\ee
where $\Delta \psi$ and $\Delta \phi$ are the ranges of $\psi$ and $\phi$ respectively.  If we took $\psi$ and $\phi$ to be periodic directions both periods would have to be integer multiples of $4 \pi$ and (\ref{area2}) can not be satisfied--in other words, if $\psi$ and $\phi$ are both periodic one has covered the unit sphere multiple times.  If we require one of the directions, say $\psi$, to be periodic then its period must be $4 \pi$ and $\phi$ (not a periodic direction) must have range $2 \pi$.  Under these circumstances one can show, with a bit of algebra, that given any Cartesian point on the sphere $(x_1, x_2, x_3, x_4)$ the values of $\psi$ and $\phi$ are uniquely determined.  Hence we have a one-to-one onto invertible map--i.e. a diffeomorphism.  If one insisted on making $\psi$ periodic with period $2 \pi$, one only has a diffeomorphism if one takes the sphere with opposite points identified ($x_i \equiv -x_i$), i.e. $S_3/Z_2$.  This will cut the area of the sphere in half, but by cutting the period of $\psi$ in half we can restore the analog of (\ref{area2}) and obtain a diffeomorphism.

\setcounter{equation}{0}
\section{Appendix II: Generalizing the Hopf fibration}

We will choose our parameters $(\alpha,\beta, \gamma, \delta)$ and ranges for our angles $\psi$ and $\phi$ such that one fundamental domain in our new coordinates $(\theta, \psi, \phi)$ is diffeomorphic to a full $S_3$.  That is, there is a one-to-one onto mapping such that if we take $\phi_1$ and $\phi_2$ as defined by (\ref{ang1}) and (\ref{ang2}) there are Cartesian coordinates $(x_1, x_2, x_3, x_4)$ that cover the sphere precisely once
\begin{eqnarray}
x_1 &=& \sin \theta \sin \phi_1\\
x_2 &= &\sin \theta \cos \phi_1\\
x_3 &= &\cos \theta \sin \phi_2\\
x_4 &=& \cos \theta \cos \phi_2
\end{eqnarray}

 Since we will insist that $\psi$ is a periodic direction (so that it may be pinched off in the interior of the spacetime) the period of $\psi$, $\Delta \psi$, must be an integer multiple of $2 \pi/\alpha$, as well as $2 \pi/\gamma$, and the ratio $\gamma/\alpha$ must be rational.  Then, as before, we take
\be \label{gammarat2}
\frac{\gamma}{\alpha} = \frac{m_1}{m_0}
\ee
where $m_1$ and $m_0$ are relatively prime integers and $m_0 > 0$.   

If $\phi$ were a periodic direction as well then likewise that would force the ratio $\beta/\delta$ to be rational.  If we insist that both $\psi$ and $\phi$ are both periodic directions with periods $2 \pi$, as we will see below, this turns out to force $\vert \alpha \delta - \beta \gamma \vert = 1$.  Given these restrictions, and also insisting $\alpha \delta - \beta \gamma > 0$, after a rescaling of the angles these transformations would be the $SL(2,\mathbb Z)$ transformations.   Note, however, $\phi$ need not necessarily be periodic for our purposes and, as we reviewed above in  Appendix I, in the simple case of the Hopf fibration if we took $\phi$ to be periodic the new coordinates can only be diffeomorphic to the quotiented sphere.  Instead we wish to restrict ourselves to the case where we have globally asymptotically AdS or globally asymptotically flat space.  $\phi$ still must have some definite range $\Delta \phi$ and this is given in terms of $\Delta \psi$ and our parameters $(\alpha,\beta, \gamma, \delta)$ by insisting that the proper area of the unit $S_3$ in these new coordinates be $2 \pi^2$:
\be
\Delta \phi = \frac{(2 \pi)^2}{ \Delta \psi \vert  \alpha \delta - \beta \gamma \vert}
\ee

\begin{figure}[t]
\centering
	\includegraphics[scale= .5]{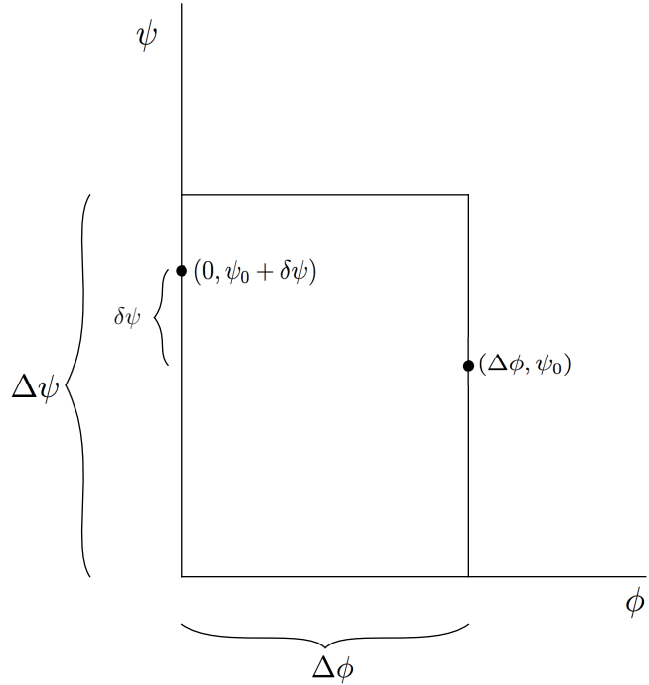}
	\caption{A fundamental domain with equivalent points on $S_3$}
	\label{DR1}
	\end{figure}

If $\phi$ is not periodic we will not have properly specified our new fundamental domain $ (0 \leq \psi < \Delta \psi, 0 \leq \phi < \Delta \phi)$ until we provide a smooth rule describing the limit as one approaches the edge of this domain.   The only rule that appears to make any sense is that $\lim_{\phi\rightarrow \Delta \phi}  (\psi, \phi) $ yields the same point on the $S_3$ as that corresponding to $(\psi + \delta \psi, 0)$, that is if we go to the edge of the domain at $\phi = \Delta \phi $ at some $\psi $ we go to $\phi = 0$, but not necessarily to the same value of $\psi$.   We have sketched the situation in Figure \ref{DR1}, drawing for simplicity the constant $\psi$ and $\phi$ lines at right angles.  As long as such a rule is satisfied, the direction $\phi$ will be periodic on any surface on which the $\psi$ direction degenerates and which all values of $\psi$ are taken simultaneously.  This is precisely the situation we encounter when we find solitons by adjusting parameters such that there is a surface where $\psi$ degenerates.  

This smoothness rule requires that these two corresponding points in the domain yield the same $\phi_1$ and $\phi_2$ up to integer multiples of $2 \pi$.  A few lines of algebra shows this can only work if the period of $\psi$ is precisely
\be
\Delta \psi = \frac{2 \pi m_0}{\vert \alpha \vert}
\ee
(recalling $\gamma/\alpha = m_1/m_0$) and that
\be \label{dpsicond}
\delta \psi = \frac{\beta}{\alpha} \Delta \phi - \frac{2 \pi p_0}{\vert \alpha \vert } \, \mathrm{sgn}(\alpha \delta - \beta \gamma)
\ee
where $(p_0,p_1)$ are integers such that
\be \label{BZ1}
p_1 m_0 - p_0 m_1 = 1
\ee
The existence of such an integers $(p_0, p_1)$ is proven by B\'ezout's identity.  There are clearly multiple solutions to (\ref{BZ1});  given one set $(p_0, p_1)$ of solutions, $(p_1 + l_0 m_1, p_0 + l_0 m_0)$ for any integer $l_0$ will work just as well.  Fortunately it is known these are the only solutions to (\ref{BZ1}) and shifting the value of $p_0 \rightarrow p_0 + l_0 m_0$ shifts $\delta \psi$ by an integer  number of periods of $\psi$, that is by $l_0 \Delta \psi$.  Hence there is only one such $p_0$ within a fundamental domain.  We note under these conditions the range of $\phi$ is
\be
\Delta \phi = \frac{2 \pi}{ \vert m_0 \delta - m_1 \beta \vert}
\ee

It is also possible to derive the above conditions using graphical methods.  For the sake of simplicity, we will restrict that discussion to the case where $\alpha > 0$ and $\alpha \delta - \beta \gamma > 0$.  Then the fact that $\psi$ is a periodic direction implies the axis $\psi$ defined by $\phi = 0$ should cross the lattice defined by $\phi_1 \sim \phi_1 + 2\pi$, $\phi_2 \sim \phi_2 + 2\pi$ and hence $\gamma/\alpha$ is rational as before (\ref{gammarat2}).  The period  $\Delta \psi$ of $\psi$ is determined by first finding on the axis $\phi = 0$ which values of $(\phi_1,\phi_2)$ correspond to the same point on the sphere as $(\phi_1=0,\phi_2 =0)$ and then evaluating $\psi$ for the couple $(\phi_1,\phi_2)$ with the smallest $\phi_1 > 0$, i.e. the first point on the axis $\phi = 0$ equivalent to $(\phi_1=0,\phi_2 =0)$.  Specifically, we find the smallest positive integers $p_0$, $p_1$ such that $\alpha \psi = 2\pi p_0$ and $\gamma \psi = 2\pi p_1$. Using \eqref{gammarat2}, we get $m_1 p_0 = m_0 p_1$. The smallest solution is $p_0 = m_0$, $p_1 = m_1$. At that point, $\psi$ has the value $2\pi m_0 / \alpha$ and hence the period of $\psi$, $\Delta \psi$ is then given by $2 \pi m_0 /\alpha$.   

The rule for the smoothness of the domain in this line of reasoning becomes the statement that  the $\phi_1$ and $\phi_2$ coordinates of the points $(\phi + \Delta \phi,\psi)$ and $(\phi,\psi+\delta \psi)$ can only differ by integer  multiples $(p_0,p_1)$ of $2 \pi$.   One has $2\pi p_0 = -\alpha \delta \psi + \beta \Delta \phi$ and $2\pi p_1 = -\gamma \delta\psi + \delta \Delta \phi$. Replacing $\delta\psi$ in one equation using the other, one gets \eqref{BZ1}. Removing $\Delta\phi$ using both equations or rewriting the first equation, one finds
\be
\delta \psi = - \frac{2 \pi}{\alpha \delta - \beta \gamma}(p_0 \delta - p_1 \beta) = \frac{\beta}{\alpha}\Delta \phi-\frac{2\pi p_0}{\alpha} \label{eqpsi}
\ee
as  before.  We have sketched the situation in Hopf fibration case in Figure \ref{DR2}. 

\begin{figure}
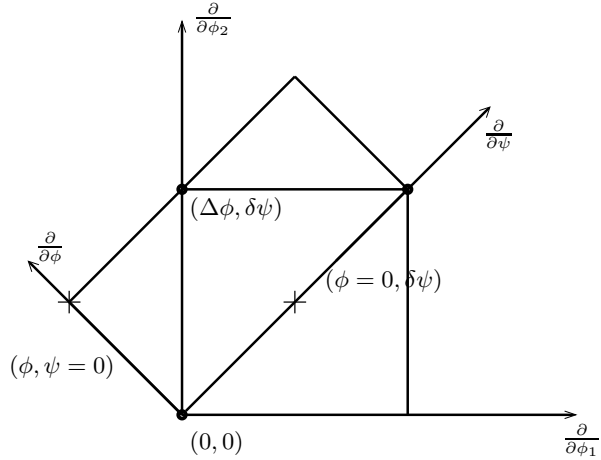

\centering
  \input hopf2.pstex_t
	\caption{Hopf fibration case. The two rectangles denote two fundamental domains at fixed azimuthal angle $\theta$. Equivalent points on $S_3$ are marked by dots and crosses. The identification of points at the extremities of the fundamental domain define $\delta\phi$.}
	\label{DR2}
	\end{figure}

\setcounter{equation}{0}
\section{Appendix III: Electric and local magnetic charges}

Let us now consider the relationship between the local magnetic charge and the global electric charge.   Since the metric and the field strength are regular everywhere, the electromagnetic field obeys the equation of motion
\be
d \star F - \frac{1}{\sqrt{3}} F \wedge F = 0 . \label{EOMF}
\ee
without sources in the whole spacetime. The gauge field even though it is abelian is self-interacting strongly in the spacetime due to the Chern-Simons coupling. 

As discussed before, these regular solutions have a net global electric charge because there is no globally defined regular potential. One may write down the magnetic charge as a quantity proportional to the difference of the gauge potential between the two patches  $\mathcal N$ and  $\mathcal S$
\be
A|_{\mathcal S}- A|_{\mathcal N} = (A_{\phi}|_{\mathcal S} - A_{\phi}|_{\mathcal N})d \phi \equiv   \Delta A_{\phi} d \phi, 
\ee
where, for definitiveness the northern region $\mathcal N$ is defined as the region of spacetime of constant time $t$ and $0 \leq \theta \leq \frac{\pi}{4}$ and the southern region $\mathcal S$ as the region of spacetime of constant time $t$ where $\frac{\pi}{4} \leq \theta \leq \frac{\pi}{2}$. Integrating the expression \eqref{EOMF} in the northern region and in the southern region and summing the contributions, we get
\begin{eqnarray}
0 &=&\int_{\mathcal N} ( d \star F - \frac{1}{\sqrt{3}} F \wedge F ) + \int_{\mathcal S} ( d \star F - \frac{1}{\sqrt{3}} F \wedge F ),\\
&=& \int_{\partial \mathcal N} ( \star F - \frac{1}{\sqrt{3}} F \wedge A ) + \int_{\partial \mathcal S} (  \star F - \frac{1}{\sqrt{3}} F \wedge A ),
\end{eqnarray}
where we used Stokes' theorem on the second line. Now, the boundary of the region $\mathcal N$ consists in the northern hemisphere of the surface $S^3_{\infty}$ at infinity and the equator $\mathcal E$. The southern region admits the southern hemisphere of $S^3_{\infty}$ and the equator $\mathcal E$ with an opposite orientation. Using the definition of electric charge at infinity \eqref{usualQE}, we thus get the identity
\begin{eqnarray}
Q_{E} =  \frac{\Delta A}{\sqrt{3}} \int_{\mathcal E} F \wedge d\phi.\label{idF}
\end{eqnarray}
Since the left-hand side is gauge invariant and the right-hand side is the integral of a closed form, the right hand side does not depend on any specific property of the surface $\mathcal E$ which could be smoothly deformed. We see explicitly from this relation that the origin of electric charge lies both in the non-linear Chern-Simons interaction and in the non-existence of a globally defined gauge field.


\vskip 3cm

    \begin{thebibliography}{99}
    
    \bibitem{ERBHreview} R. Emparan and H. S. Reall, ``Black Holes in Higher Dimensions,'' Living Rev. Rel. {\bf 11} 6 (2008)  [arXiv:0801.3471]
    
    
    \bibitem{Gannon} D. Gannon, J. Math. Phys. {\bf 16} 2364 (1975); G. J. Galloway, J. Phys. A, {\bf 16}, 1435 (1983)
    
    \bibitem{topcensor} J. L. Friedman, K. Schleich, and D. W. Witt, ``Topological Censorship,'' Phys. Rev. Lett. {\bf 71} (1993), 1486  [arXiv:gr-qc/9305017]; G. J. Galloway, K. Schleich, D. M. Witt, and E. Woolgar, ``The AdS/CFT Correspondence Conjecture and Topological Censorship,'' Phys. Lett. {\bf B505} 255-262(2001)  [arXiv:hep-th/9912119]
    
       \bibitem{Wittenbubble}
 E.~Witten,
  ``Instability Of The Kaluza-Klein Vacuum,''
  Nucl.\ Phys.\ B {\bf 195} (1982) 481; 
D.~Brill and H.~Pfister, ``States of negative total energy in Kaluza-Klein theory,'' Phys. Lett. B {\bf 228}, 359 (1989); D.~Brill and G.~T.~Horowitz, ``Negative energy in string theory,'' ' Phys. Lett. B {\bf 262}, 437 (1991).



    
    \bibitem{EH} R. Clarkson and R. B. Mann, ``Soliton Solutions to the Einstein Equations in Five Dimensions,'' Phys. Rev. Lett. {\bf 96} 051104 (2006)  [arXiv:hep-th/0508109]; K. Copsey, ``Bubbles Unbound II: AdS and the Single Bubble,'' JHEP {\bf 0710} 095 (2007)
    
     \bibitem{GibbonsSquashed}
  P. Bizon, T. Chmaj, G. W. Gibbons, and C. N. Pope, ``Gravitational Solitons and the Squashed Seven-Sphere,'' Class. Quant. Grav. {\bf 24} 4751-4776 (2007)  [arXiv:hep-th/0701190].
    
     \bibitem{Mizoguchi:1998wv}
S. Mizoguchi and N. Ohta. ``More on the similarity between D = 5 simple supergravity and M
  theory,''  Phys. Lett. {\bf B441} 123--132 (1998)  [arXiv:hep-th/9807111]

\bibitem{CJLP}
E.~Cremmer, B.~Julia, Hong Lu, and C.~N. Pope.
``Higher-dimensional origin of d = 3 coset symmetries,'' 1999, [arXiv:hep-th/9909099];  
``Dualisation of dualities. I,''  Nucl. Phys. {\bf B523} :73--144 (1998)   [arXiv:hep-th/9710119];
``Dualisation of dualities. II: Twisted self-duality of doubled fields
  and superdualities,''  Nucl. Phys. {\bf B535} 242--292 (1998)   [arXiv: hep-th/9806106.].
  
\bibitem{Bouchareb:2007ax}
Adel Bouchareb et~al. ``$G_2$ generating technique for minimal D=5 supergravity and black
  rings,'' Phys. Rev. {\bf D76} 104032 (2007),   [arXiv: 0708.2361]

    
    \bibitem{GSG2} G. Comp\`ere, S. de Buyl, E. Jamsin, A. Virmani, ``G2 Dualities in $D = 5$ Supergravity and Black Strings,''  Class. Quant. Grav. {\bf 26} (2009) 125016  [arXiv:0903.1645] 
        
      \bibitem{CCLP05}
 Z. W. Chong, M. Cvetic, H. Lu, and C.N. Pope ``General Non-Extremal Rotating Black Holes in Minimal Five-Dimensional Gauged Supergravity'' hep-th/0506029.


    
    \bibitem{TYI09}
S. Tomizawa, Y. Yasui, and A. Ishibashi, ``A uniqueness theorem for charged rotating black holes in five-dimensional minimal supergravity,'' hep-th/0901.4724

\bibitem{Witten1981mf} E.~Witten, ``A Simple Proof Of The Positive Energy Theorem,''  Commun.\ Math.\ Phys.\  {\bf 80}, 381 (1981).

\bibitem{EmparanDipBR04} R. Emparan, ``Rotating circular strings, and infinite non-uniqueness of black rings,'' JHEP {\bf 0403} (2004) 0604 [arXiv:hep-th/0402149]

\bibitem{RossBubbles}
 S.~Ross,
  ``Non-Supersymmetric Asymptotically $AdS_5 \times S^5$ Smooth Geometries,''
  JHEP {\bf 0601} (2006) 130  
   [arXiv:hep-th/0511090].
    

\bibitem{CveticEmbed} M. Cvetic et al, ``Embedding AdS black holes in ten and eleven dimensions,'' Nucl. Phys. {\bf B558} (1999) 96-126 [arXiv:hep-th/9903214]

\bibitem{HollandsIshibashiMarolf} S. Hollands, A. Ishibashi, and D. Marolf, ``Comparison between various notions of conserved charges in asymptotically AdS-spacetimes,'' Class. Quant. Grav. {\bf 22} (2005) 2881-2920 [arXiv:hep-th/0503045]

\bibitem{AshtekarDas} A. Ashtekar and S. Das, ``Asymptotically Anti-de Sitter Space-times: Conserved Quantities,'' Class. Qaunt. Grav. {\bf 17} (2000) L17-L30  [arXiv:hep-th/9911230]

\bibitem{GKLTT}
G. W. Gibbons, D. Kastor, L. A. J. London, P. K. Townsend, and J. Traschen, ``Supersymmetric Self-Gravitating Solitons,''' hep-th/9310118

\bibitem{CopseyHorowitz}
K. Copsey and G. T. Horowitz, ``The Role of Dipole Charge in Black Hole Thermodynamics,'' Phys. Rev. {\bf D73} (2006) 024015 [arXiv:hep-th/0505278]

 \bibitem{MP}
  R.~C.~Myers and M.~J.~Perry, ``Black Holes in Higher Dimensions,'' Annals Phys. {\bf 172} 304, 1986.
  
  \bibitem{Barnichetal1}
  G. Barnich and F. Brandt, ``Covariant theory of asymptotic symmetries, conservation laws and central charges,'' Nucl. Phys.{\bf B633} 3-82 (2002)   [arXiv:hep-th/0111246] ; G. Barnich, ``Boundary charges in gauge theories: Using Stokes theorem in the bulk,'' Class. Quant. Grav. {\bf 20} 3685-3698 (2003)   [arXiv:hep-th/0301039] ; G. Barnich and G. Comp\`ere, ``Surface charge algebra in gauge theories and thermodynamic integrability,'' J. Math. Phys {\bf 49} 042901 (2008) [arXiv: 0708.2378] 
  
  \bibitem{Barnich:2005kq}
  G. Barnich and G. Comp\`ere, ``Conserved charges and thermodynamics of the spinning Goedel black hole,'' Phys. Rev. Lett {\bf 95} 031302 (2005) [arXiv:hep-th/0501102] 
  
  \bibitem{GMT99}
  J. P. Gauntlett, R. C. Myers, and P. K. Townsend, ``Black Holes of $d = 5$ Supergravity,'' Class. Quant. Grav. {\bf 16} 1-21 (1999)  [arXiv:hep-th/9810204] 
  
  \bibitem{Iyer:1994ys}
  V.~Iyer and R.~M.~Wald,
  ``Some properties of Noether charge and a proposal for dynamical black hole
  entropy,''
  Phys.\ Rev.\  D {\bf 50}, 846 (1994)
  [arXiv:gr-qc/9403028].

  
  \bibitem{HawkingHorowitz}
 S. W. Hawking and G.T. Horowitz, ``The Gravitational Hamiltonian, Action, Entropy, and Surface Terms,'' Class. Quant. Grav. 13 (1996) 1487-1498.

  
  \bibitem{AdSBPS}
  L. A. J. London, ``Arbitrary dimensional cosmological multi-black holes,'' Nucl. Phys. B {\bf 434} (1995) 709-735
  
  
  \bibitem{GutowskiReall} J. B. Gutowski and H. S. Reall, ``Supersymmetric $AdS_5$ black holes,'' JHEP {\bf 0402} (2004) 006  [arXiv:hep-th/0401042]
  
 
\bibitem{Clement:2008qx}
G. Clement, ``Sigma-model approaches to exact solutions in higher- dimensional
  gravity and supergravity,''  [arXiv: 0811.0691];  ``The symmetries of five-dimensional minimal supergravity reduced to
  three dimensions,'' J. Math. Phys. {\bf 49} 042503 (2008) [arXiv:0710.1192].

\bibitem{Gal'tsov:2008sh}
D. V. Gal'tsov and N. G. Scherbluk,``Improved generating technique for D=5 supergravities and squashed Kaluza-Klein Black Holes,'' [arXiv: 0812.2336] 

\bibitem{Gal'tsov:2008nz}
D. V. Gal'tsov and N.~G. Scherbluk, ``Generating technique for $U(1)^3 \, 5D$ supergravity,''
Phys. Rev. {\bf D78} 064033 ( 2008)  [arXiv:0805.3924]



\bibitem{ErgInstab}
V. Jejjala, O. Madden, S. F. Ross, and G. Titchener, ``Non-supersymmetric smooth geometries and $D1-D5-P$ bound states,'' Phys. Rev. {\bf D71} 124030 (2005) [arXiv: hep-th/0504181]; V. Cardoso, O. J. C. Dias, J. L. Hovdebo, and R. C. Myers, ``Instability of non-supersymmetric smooth geometries,'' Phys. Rev. {\bf D73} 064031 (2006)    [arXiv: hep-th/0512277]
  
\bibitem{WittenOlive}
E. Witten and D. Olive, ``Supersymmetry Algebras that include Topological Charges,'' Phys. Lett. {\bf 78B} no. 1:97 (1978).


  \end{thebibliography}
 \end{document}